\input harvmac.tex
\input labeldefs.tmp
\writedefs
\overfullrule=0mm
\hfuzz 10pt

\def\modif#1{} 

\input epsf.tex
\newcount\figno
\figno=0
\def\fig#1#2#3{
\par\begingroup\parindent=0pt\leftskip=1cm\rightskip=1cm\parindent=0pt
\baselineskip=11pt
\global\advance\figno by 1
\midinsert
\epsfxsize=#3
\centerline{#2}
\vskip 10pt
{\bf Fig.~\the\figno:} #1\par
\endinsert\endgroup\par
}
\def\figlabel#1{\xdef#1{\the\figno}\writedef{#1\leftbracket \the\figno}}
\def\encadremath#1{\vbox{\hrule\hbox{\vrule\kern8pt\vbox{\kern8pt
\hbox{$\displaystyle #1$}\kern8pt}
\kern8pt\vrule}\hrule}}
\newdimen\xfigunit
\global\xfigunit=4144sp
%
\newcount\x \newcount\y
\newcount\newmag\newcount\oldmag
\def\pic(#1,#2)(#3,#4)#5{%
\global\x=-#3%
\global\advance\x by-#1%
\global\y=-#4%
\global\newmag=\epsfxsize%
\global\divide\newmag by \xfigunit%
\def\epsfsize##1##2{\expandafter\epsfxsize%
\global\oldmag=##1\global\divide\oldmag by \xfigunit}%
\epsfbox{#5}%
}
\def\put(#1,#2)#3{%
{%
\advance\x by#1\advance\y by#2%
\multiply\x by \newmag%
\divide\x by \oldmag%
\multiply\y by \newmag%
\divide\y by \oldmag%
\rlap{\kern\x\xfigunit%
\raise\y\xfigunit\hbox{#3}}}}
%
%
%
%
%
\def\frac#1#2{{\scriptstyle{#1 \over #2}}}

%
%

%
\def\({ \left( }\def\[{ \left[ }
\def\){ \right) }\def\]{ \right] }
%


\def\IR{\relax{\rm I\kern-.18em R}}
\font\cmss=cmss10 \font\cmsss=cmss10 at 7pt
\def\IZ{\relax\ifmmode\mathchoice
{\hbox{\cmss Z\kern-.4em Z}}{\hbox{\cmss Z\kern-.4em Z}}
{\lower.9pt\hbox{\cmsss Z\kern-.4em Z}}
{\lower1.2pt\hbox{\cmsss Z\kern-.4em Z}}\else{\cmss Z\kern-.4em Z}\fi}
\def\inbar{\,\vrule height1.5ex width.4pt depth0pt}
\def\IB{\relax{\rm I\kern-.18em B}}
\def\IC{\relax\hbox{$\inbar\kern-.3em{\rm C}$}}
\def\ID{\relax{\rm I\kern-.18em D}}
\def\IE{\relax{\rm I\kern-.18em E}}
\def\IF{\relax{\rm I\kern-.18em F}}
\def\IG{\relax\hbox{$\inbar\kern-.3em{\rm G}$}}
\def\IH{\relax{\rm I\kern-.18em H}}
\def\II{\relax{\rm I\kern-.18em I}}
\def\IK{\relax{\rm I\kern-.18em K}}
\def\IL{\relax{\rm I\kern-.18em L}}
\def\IM{\relax{\rm I\kern-.18em M}}
\def\IN{\relax{\rm I\kern-.18em N}}
\def\IO{\relax\hbox{$\inbar\kern-.3em{\rm O}$}}
\def\IP{\relax{\rm I\kern-.18em P}}
\def\IQ{\relax\hbox{$\inbar\kern-.3em{\rm Q}$}}
\def\IGa{\relax\hbox{${\rm I}\kern-.18em\Gamma$}}
\def\IPi{\relax\hbox{${\rm I}\kern-.18em\Pi$}}
\def\ITh{\relax\hbox{$\inbar\kern-.3em\Theta$}}
\def\IOm{\relax\hbox{$\inbar\kern-3.00pt\Omega$}}

\def\Z{\IZ}


\def\oh{{1\over 2}}

\def\Ga{\alpha}
\def\Gd{\delta}

\def\Gs{\sigma}\def\Gt{\tau}


\def\bra{\langle}\def\ket{\rangle}

\def\hepth#1{{\tt hep-th #1}}

\def\\#1 {{\tt\char'134#1} }

\catcode`\@=11
\def\Eqalign#1{\null\,\vcenter{\openup\jot\m@th\ialign{
\strut\hfil$\displaystyle{##}$&$\displaystyle{{}##}$\hfil
&&\qquad\strut\hfil$\displaystyle{##}$&$\displaystyle{{}##}$
\hfil\crcr#1\crcr}}\,}   \catcode`\@=12
\def\encadre#1{\vbox{\hrule\hbox{\vrule\kern8pt\vbox{\kern8pt#1\kern8pt}
\kern8pt\vrule}\hrule}}
\def\encadremath#1{\vbox{\hrule\hbox{\vrule\kern8pt\vbox{\kern8pt
\hbox{$\displaystyle #1$}\kern8pt}
\kern8pt\vrule}\hrule}}


%
\newdimen\xraise\newcount\nraise
\def\xpoint{\hbox{\vrule height .45pt width .45pt}}
\def\udiag#1{\vcenter{\hbox{\hskip.05pt\nraise=0\xraise=0pt
\loop\ifnum\nraise<#1\hskip-.05pt\raise\xraise\xpoint
\advance\nraise by 1\advance\xraise by .4pt\repeat}}}
\def\ddiag#1{\vcenter{\hbox{\hskip.05pt\nraise=0\xraise=0pt
\loop\ifnum\nraise<#1\hskip-.05pt\raise\xraise\xpoint
\advance\nraise by 1\advance\xraise by -.4pt\repeat}}}
\def\vertex{\epsfxsize=5mm\hbox{\raise -2mm\hbox{\epsfbox{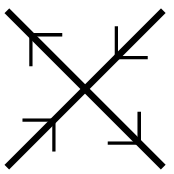}}}}
\def\vertexalt{\epsfxsize=5mm\hbox{\raise -2mm\hbox{\epsfbox{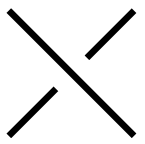}}}}
\def\vertexdbl{\epsfxsize=8mm\hbox{\raise -4mm\hbox{\epsfbox{link03.eps}}}}
\def\propagdbl{\epsfxsize=12mm\hbox{\raise -1mm\hbox{\epsfbox{link04.eps}}}}
\def\vertexarr{\epsfxsize=8mm\hbox{\raise -4mm\hbox{\epsfbox{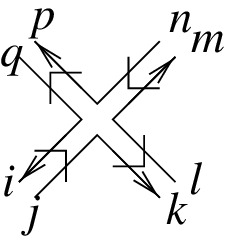}}}}
\def\propagarr{\epsfxsize=12mm\hbox{\raise -1mm\hbox{\epsfbox{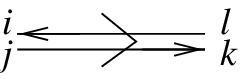}}}}
\def\vcralt{\epsfxsize=28mm\hbox{\raise -2mm\hbox{\epsfbox{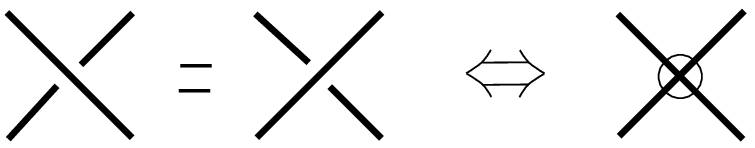}}}}
\def\Th{Thistlethwaite}

\def\ommit#1{}
\def\hepth#1{{\tt hep-th/#1}}\def\mathph#1{{\tt math-ph/#1}}  
\def\S{S}
%
%

\lref\Ada{P. Adamietz, {\sl Kollektive Feldtheorie und Momentenmethode 
in Matrixmodellen}, PhD thesis, internal report DESY T-97-01.}
\lref\Ake{G. Akemann, unpublished notes,  private communication.}
\lref\AKM{J. Ambj\o rn, C.F. Kristjansen, Y.M. Makeenko, 
{\sl Higher Genus Correlators for the Complex Matrix Model},
{\it Mod.Phys.Lett.} {\bf  A7} (1992) 3187-3202, \hepth{9207020}.}
\lref\BF{A. Bartholomew and R. Fenn, 
  {\sl Quaternionic Invariants of Virtual Knots and Links}, preprint.}
\lref\BIZ{D. Bessis, C. Itzykson and J.-B. Zuber, 
{\sl Quantum Field Theory Techniques in Graphical Enumeration},
{\it Adv. Appl. Math.} {\bf 1} (1980) 109--157.}
\lref\BMS{M. Bousquet-M\'elou and G. Schaeffer,
{\sl Enumeration of planar constellations},
{\it Adv. in Appl. Math.} 24(4) (2000)
337--368\semi
G. Schaeffer, {\sl Conjugaison d'arbres et cartes combinatoires
al\'eatoires}, Th\`ese de doctorat,
{\tt
http://dept-info.labri.u-bordeaux.fr/$\sim$schaeffe/cv/bibli/These.html}}
\lref\BIPZ{E. Br\'ezin, C. Itzykson, G. Parisi and J.-B. Zuber, 
{\sl Planar Diagrams}, {\it Commun. Math. Phys.} {\bf 59} (1978) 35--51.}
\lref\Coll{B. Collins, {\sl Moments and cumulants of polynomial
random variables on unitary groups, the Itzykson-Zuber integral
and free probability}, \mathph{0205010},
{\it Int.Math.Res.Notices}, to appear.}     
\lref\DFGZJ{P. Di Francesco, P. Ginsparg and J. Zinn-Justin, 
{\sl 2D Gravity and Random Matrices, }{\it Phys. Rep.} {\bf 254} (1995)
1--133, \hepth{9306153}.}
\lref\tH{G. 't Hooft, 
{\sl A Planar Diagram Theory for Strong 
Interactions}, {\it Nucl. Phys.} {\bf B 72} (1974) 461--473.}
\lref\HTW{J. Hoste, M. Thistlethwaite and J. Weeks, 
{\sl The First 1,701,936 Knots}, {\it The Mathematical Intelligencer}
{\bf 20} (1998) 33--48.}
\lref\JZJ{J. Jacobsen and P. Zinn-Justin,
{\sl A transfer matrix approach to the enumeration of knots},
{\it  J. Knot Theory and its Ramifications}, {\bf 11} (2002) 739-758.}
\lref\LK{L. Kauffman,
{\sl Virtual knot theory}, 
{\it Europ. J. Combin.}  {\bf 20} (1999) 663-690, 
{\tt math.GT/9811028};
{\sl Detecting virtual knots},  Chicago preprint.
}
\lref\KR{L. Kauffman and D.E. Radford,
{\sl Bioriented Quantum Algebras, and a Generalized
Alexander Polynomial for Virtual Links }, 
 http://www.math.uic.edu/$\sim$kauffman/Papers.html}
\lref\KM{V.A.~Kazakov and A.A.~Migdal, {\sl Recent progress in the
theory of non-critical strings}, {\it Nucl. Phys.} {\bf B 311} (1988)
171--190.}
\lref\KP{V.A.~Kazakov and P.~Zinn-Justin, {\sl Two-Matrix Model with
$ABAB$ Interaction}, {\it Nucl. Phys.} {\bf B 546} (1999) 647--668.}
\lref\Ku{G. Kuperberg, {\sl What is a virtual link?}, {\tt
math.GT/0208039}.}
\lref\MTh{W.W.~Menasco and M.B.~\Th, 
{\sl The Tait Flyping Conjecture}, {\it Bull. Amer. Math. Soc.} {\bf 25}
(1991) 403--412;
{\sl The Classification of Alternating 
Links}, {\it Ann. Math.} {\bf 138} (1993) 113--171.}
\lref\Mo{T.R. Morris, {\sl Chequered surfaces and complex matrices},
{\it Nucl. Phys.} {\bf B 356} (1991) 703-728.}

\lref\Ro{D. Rolfsen, {\sl Knots and Links}, Publish or Perish, Berkeley 1976.}
\lref\Saw{J.~Sawollek, {\sl On Alexander--Conway polynomials for virtual
knots and links}, {\tt math.GT/9912173}.}
\lref\STh{C. Sundberg and M. Thistlethwaite, 
{\sl The rate of Growth of the Number of Prime Alternating Links and 
Tangles}, {\it Pac. J. Math.} {\bf 182} (1998) 329--358.}
\lref\SW{D.S. Silver and S.G. Williams, {\sl Alexander groups and
virtual links}, {\it J. Knot Theory and its Ramifications} to appear.}
\lref\Tutmap{W.T.~Tutte, {\sl A Census of Planar Maps}, 
{\it Can. J. Math.} {\bf 15} (1963) 249--271.}
\lref\Tutpol{W.T.~Tutte, {\sl A Census of Hamiltonian polygons}, 
{\it Can. J. Math.} {\bf 14} (1962) 402--417.}
\lref\virt{S.-G. Kim, {\sl Virtual knot groups}, {\tt
math.GT/9907172};
J.S. Carter, S. Kamada and M.  Saito, {\sl Stable equivalence of knots 
on surfaces and virtual knot cobordisms}, 
{\it J. Knot Theory and its Ramifications}, {\bf 11} (2002) 311--322, 
{\tt math.GT/0008118}. }
\lref\PZJ{P.~Zinn-Justin, 
{\sl The General $O(n)$ Quartic Matrix Model and its application to
Counting Tangles and Links}, \mathph{0106005}.}
\lref\ZJZun{P.~Zinn-Justin and J.-B. Zuber, 
{\sl Matrix integrals and the counting of tangles and links}, 
{\it Discr. Math.} {\bf 246} (2002) 343-360, \mathph{9904019}.} 
\lref\ZJZdeu{P.~Zinn-Justin and J.-B. Zuber, 
{\sl On the Counting of Coloured Tangles}, {\tt math-ph/0002020},  
{\it J. Knot Theory and its Ramifications}, {\bf 9} (2000) 1127--1141.}
\lref\Zv{A.~Zvonkin, {\sl Matrix Integrals and Map Enumeration: An Accessible
Introduction},
{\it Math. Comp. Modelling} {\bf 26} (1997) 281--304.}
\lref\KJS{S.~Kunz-Jacques and G.~Schaeffer,
{\sl The asymptotic number of prime alternating links},
Proceedings of the 14th international conference on Formal Power Series 
and Algebraic Combinatorics, Phenix, 2001.
}
%
\Title{
\vbox{\baselineskip12pt\hbox{SPhT 03/032}\hbox{{\tt math-ph/0303049}}}
}
{{\vbox {
\vskip-10mm
\centerline{Matrix Integrals and the Generation and Counting}
\medskip\centerline{of Virtual Tangles and Links}
}}}
\medskip
\centerline{P. Zinn-Justin}\medskip
\centerline{\it Laboratoire de Physique Th\'eorique et Mod\`eles Statistiques}
\centerline{\it Universit\'e Paris-Sud, B\^atiment 100}
\centerline{\it F-91405 Orsay Cedex, France}
\bigskip
\centerline{and}
\medskip
\centerline{J.-B. Zuber}\medskip
\centerline{\it C.E.A.-Saclay, Service de Physique Th\'eorique de Saclay,}
\centerline{\it CEA/DSM/SPhT, Unit\'e de recherche associ\'ee au CNRS}
\centerline{\it F-91191 Gif sur Yvette Cedex, France}

\vskip .2in

\noindent 
Virtual links are generalizations of classical links that
can be represented by links embedded in a
``thickened'' surface $\Sigma\times I$, product of a Riemann surface of
genus $h$ with an interval.
In this paper, we show that virtual alternating 
links and tangles are naturally associated with the $1/N^2$ expansion 
of an integral over $N\times N$ complex 
matrices. We suggest that 
it is sufficient to count the equivalence classes of these diagrams 
modulo ordinary (planar) flypes. To test this hypothesis,
we use an algorithm coding the corresponding Feynman 
diagrams by means of permutations that generates virtual diagrams 
up to 6 crossings and computes various invariants.
Under this hypothesis, we use known results on matrix integrals
to  get the generating functions of virtual 
alternating tangles of genus 1 to 5 up to order 10
(i.e.\ 10 real crossings). The asymptotic behavior for $n$ large of the
numbers of links and tangles of genus $h$ and with $n$ crossings
is also computed for $h=1,2,3$ and conjectured for general $h$.

\bigskip

\Date{03/03} 
%
\vfill\eject

\secno=-1
\newsec{Introduction}
\noindent
Virtual knots have been introduced by Kauffman \LK\ as an extension 
of classical knots. 
They may be defined as equivalence classes of 4-valent
(``4-regular'') diagrams with the ordinary under- or over-crossings
of knot theory, plus a new type of {\it virtual} crossing, 
depicted with a small circle around the intersection, see fig. 1. 
\fig{Ordinary and virtual crossings}{\epsfbox{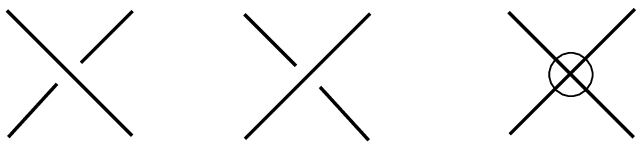}}{4cm} \figlabel\virtcr{}
\fig{Ordinary and virtual Reidemeister
moves}{\epsfbox{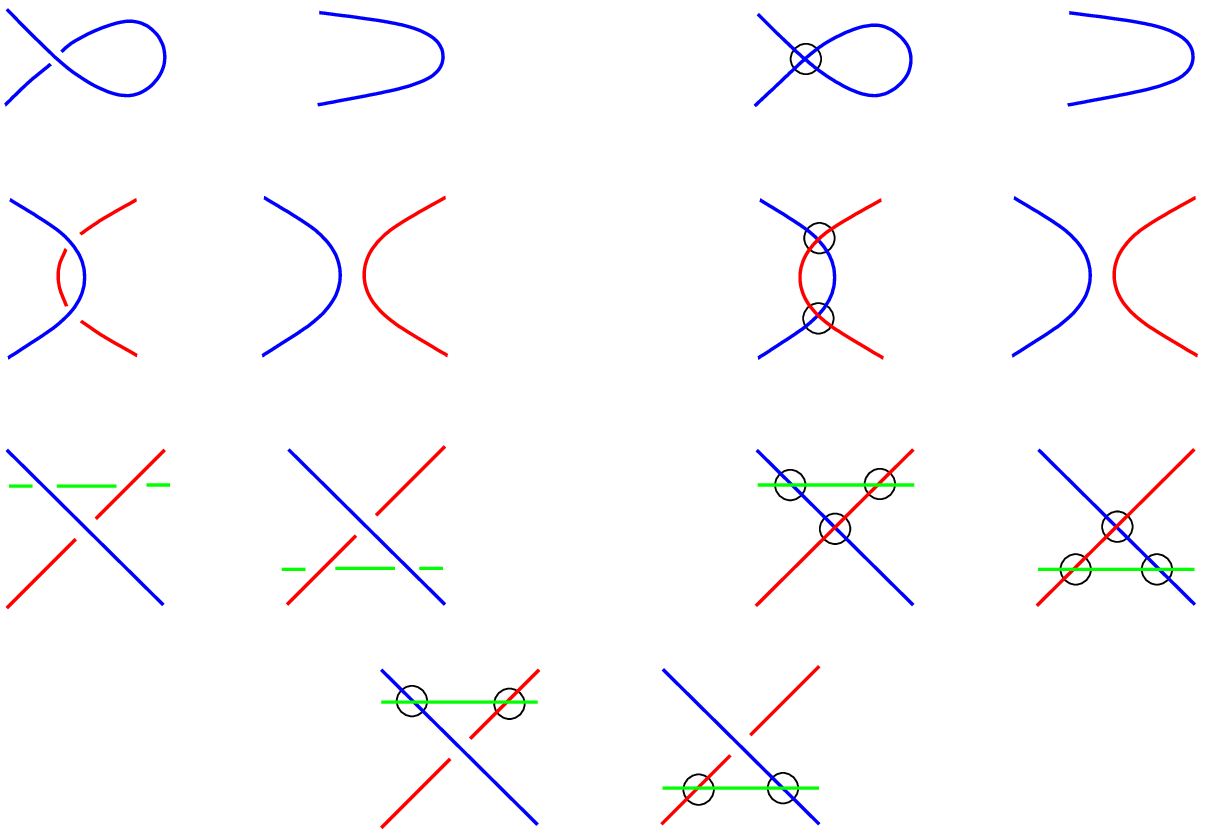}}{8cm}\figlabel\reidem
Two such diagrams are equivalent if they may be connected by
a sequence of generalized Reidemeister moves, see fig. 2.
An example of a virtual link is provided by the following
\fig{A virtual link (a) in the previous notation; (b) 
as drawn on a genus 1 surface; (c) alternative representations, 
see below.}{\epsfbox{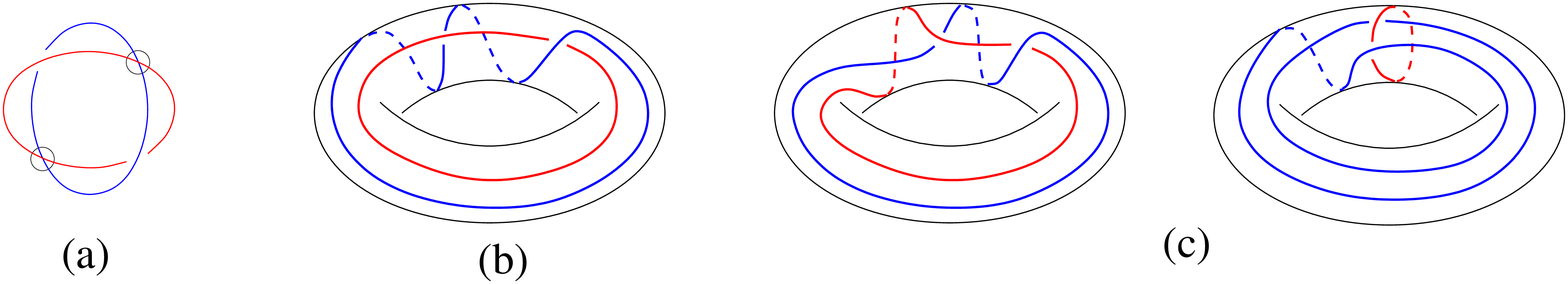}}{13cm}
\figlabel\virtlk
\fig{A forbidden Reidemeister move}{\epsfbox{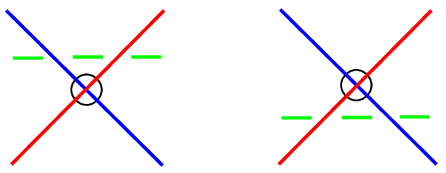}}{3cm}\figlabel\forbid
In this paper, we also use a different standpoint and notation,
closer  
to graph theory: the ordinary crossings are regarded as vertices of
a graph, the latter are regarded as rigid, i.e.\ the cyclic order 
of lines emanating from them cannot be changed, which essentially
defines a (combinatorial) {\it map}, 
and the virtual crossings are artefacts forced in the planar 
representation by the connections between vertices. In this picture, 
the meaning of these new Reidemeister moves is clear:
lines involving virtual crossings may be freely moved across 
the diagram, while keeping their end-points attached to
vertices. 
Also natural in this picture is the impossibility
for a virtual crossing to pass a line between two ordinary crossings,  
cf fig.~\forbid.

Virtual knots (or links) may also be thought of as 
drawn in the vicinity 
of a connected compact orientable Riemann surface $\Sigma$ of genus $h$, 
($h$ for ``handles''), i.e.\ embedded into the ``thickened'' surface
$\Sigma\times I$, with $I$ an interval.
The ordinary over-/under-crossings represent the projection of this
knot on $\Sigma$, while the virtual ones represent the crossing of
strands on different faces of $\Sigma$ as seen in perspective. See fig.
\virtlk(b) for an example. To obtain virtual knots,
we must consider equivalence classes of such embedded knots 
modulo isotopy in $\Sigma\times I$, and modulo orientation-preserving
homeomorphisms of $\Sigma$, 
and addition or subtraction of empty handles, see \refs{\LK,\Ku}. 
This means that we are
interested in virtual link diagrams as drawn on ``abstract'' Riemann surfaces, 
i.e.\ independently
of the actual embedding and without any preferred choice of
homology basis
(See for example in Fig. \virtlk (b)-(c) 
three equivalent representations of the virtual link (a) 
obtained from one another by various modular transformations).
As we shall see, this is precisely what
Feynman diagrams of a matrix integral do for us. 

We shall use for virtual objects the same terminology
of knots, links and tangles as for classical objects: 
a link has several
connected components, while a knot has only one.
Tangles, more precisely 4-tangles, have four open ends.

\fig{The flype of a tangle}{\epsfbox{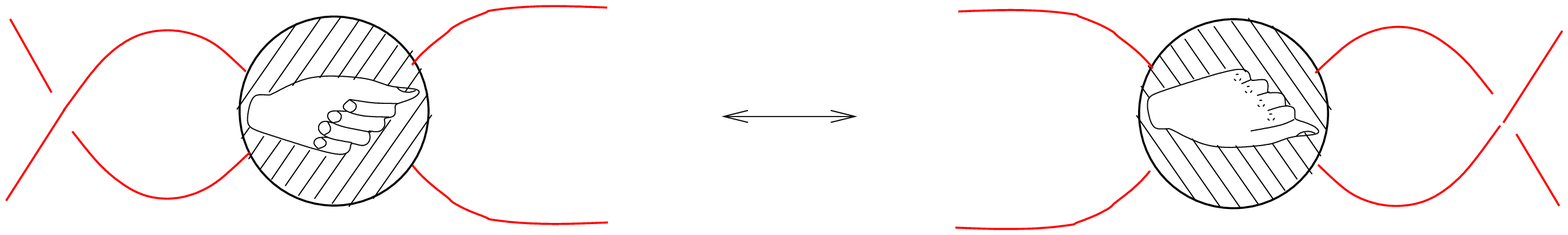}}{9cm}\figlabel\flyp
In the same way as 
{\it alternating}\/ knots/links/tangles
constitute an important subclass of the classical knotted objects,
it is natural to define alternating virtual knots/links/tangles:
they are simply described by diagrams of the previous type,
with the condition that along any strand, one encounters alternatingly 
under- and over-crossings, ignoring possible
virtual crossings. Of course, these diagrams have to be divided
by the equivalence under Reidemeister moves or some combinations
thereof. 
For classical alternating links/knots, it is a 
famous result, conjectured long ago by Tait and finally proved
by Menasco and \Th\ \MTh, that it is sufficient to consider 
{\it reduced}\/ diagrams and act on them with {\it flypes}.
Flypes are combinations of Reidemeister moves that preserve the
alternating character, see fig. \flyp. 

In a previous work \ZJZun, we have shown that  the integral
\eqn\Ia{Z_N(g,\Ga)
=\int dM 
  \exp-N\, \tr \left( \alpha  MM^\dagger 
-{g\over 2}(MM^\dagger)^2\right)\ .}
over $N\times N$  complex matrices is well suited
for the counting of alternating  links and tangles: for an appropriate 
choice of $\alpha(g)$, see below, $2{\partial\over \partial
g}\lim_{N\to \infty}    {1\over N^2}\log Z_N(g,\alpha(g))$ 
is the generating function
of the numbers of alternating tangle diagrams with $n$ 
4-valent crossings, 
and eliminating the equivalence under the flypes
just amounts to a 
``coupling constant renormalization'' \PZJ, as recalled below. In that way
the results of \STh\ were reproduced. 

\medskip
It has been known to physicists since the pioneering work of 't Hooft
\tH\ that the large $N$ limit of the previous integral may be 
organized in a topological way. While the 
leading term corresponds to {\it planar}\/ 
diagrams (in fact, drawn on a  sphere), the subdominant terms of
order $N^{-2h}$ of ${1\over N^2}\log Z(g)$ 
are described by
graphs drawn on a Riemann surface of genus $h$. It is thus quite 
natural to expect that they will be in correspondence 
with virtual link diagrams, (or after differentiation with respect to
$g$, with virtual tangle diagrams) and this is what we shall discuss and prove 
in the following.

This paper is organized as follows. We first recall (sect.~1) the 
dictionary between link/tangle diagrams and the ``Feynman diagrams''
of the matrix integral \Ia, and the necessary steps to 
eliminate the redundancy in the counting.
First remove spurious diagrams, non-prime 
or ``nugatory'' in the knot terminology, by a suitable choice of
the function $\alpha(g)$. Then we must
address the burning question of eliminating the remaining 
redundancies. 
We argue -- but this remains a heuristic argument -- 
that the same (planar) flypes as for classical alternating links and tangles
are still sufficient to remove the redundancies of alternating 
virtual diagrams. We call this the ``generalized flype conjecture'' and defer until
sect.~4 arguments in favor of this conjecture.
The flypes are taken into account by a redefinition
(``renormalization'') of the coupling constant: $g\to g_0(g)$.
In sect.~2 we apply these ideas to the case $N=1$, which corresponds to the
enumeration irrespective of the
genus; this serves as a sum rule in what follows.
In sect.~3 the explicit expressions 
of the first four terms in the large $N$ expansion of $F(g,N)$ 
are presented, corresponding to genus $h=0,1,2$ and $3$ respectively: 
while the leading term is well known (and has been used in \ZJZun)
and the second one ($h=1$) has been derived by Morris \Mo, 
the genus 2 and 3 contributions computed by Akemann and by Adamietz 
\refs{\Ake,\Ada} from the work of \AKM\  may be less known. 
Using the results of previous section
and computing the lowest contributions of a given genus (Appendix A)
gives us enough information to completely determine the numbers
of all virtual diagrams up to 11 crossings for the links, 
or 10 for the tangles, and up to genus 5.  
We also give tables of virtual alternating links up to four (real)
crossings. 
The rate of growth of the number of virtual diagrams of genus 1, 2 and 3
is  also derived from these expressions, and a general Ansatz 
is proposed for generic $h$, following what is regarded as
standard lore by physicists. 
Using the generalized flyping conjecture, we
then obtain the generating functions of virtual alternating 
4-tangles for genus 0 to 3. 
Their asymptotic behaviors for large number of (real) crossing
are, up to a larger radius of convergence, the same as
the previous ones, and under
the very plausible assumption (proved in \KJS\ for classical 
links) that generic links have no symmetry, 
we can also estimate the asymptotic number of 
virtual alternating links of a given genus.

Finally, in sect.~4, we recall that there exists a way to 
encode the relevant Feynman diagrams by means of permutations.
This is presumably
an old idea; in the present context, it seems to be  
due to Drouffe, it was used in \BIZ, 
and more recently in related topics by combinatorialists
\refs{\BMS-\Coll}. Here we use
it to set up an algorithm which is able to build all virtual link/tangle
diagrams (up to six/five crossings). 
We have used this to test the ``generalized flype conjecture'' 
by constructing as many invariants as possible, to make sure that the objects that 
cannot be obtained by flypes from each other are indeed topologically distinct. 
We provide samples of the data thus produced (the full output being accessible on the web:
{\tt http://ipnweb.in2p3.fr/$\sim$lptms/membres/pzinn/virtlinks}), and discuss
the conclusions one can draw from them.


\newsec{Matrix integrals and virtual links}
\subsec{Feynman rules for matrix integrals}
\noindent
We first recall the diagrammatic techniques to derive a
series expansion in $g$ of the integral \Ia: see \refs{\BIZ,\Zv} 
for a general introduction and \ZJZun\ for a discussion in the present
context.
 The integration measure in \Ia\ is 
\eqn\dM{ dM=\prod_{1\le i, j\le N} d\Re e\, M_{ij}\,\, d\Im m\, M_{ij}\ . }
We shall be mostly interested in the ``free energy''
\eqn\Freeen{F(g,\alpha)=\lim_{N\to\infty} 
{1\over N^2} \log {Z(\alpha,g) \over Z(\alpha,0)}}
and its derivatives.
The constant $\alpha$ can be absorbed in a rescaling $M\to \alpha^{-\oh} M$:
\eqna\Ib
$$\eqalignno{
Z(g,\alpha)&=\alpha^{-{N^2\over 2}} Z\big({g\over\alpha^2},1\big)&\Ib a\cr
F(g,\alpha)&= F\big({g\over\alpha^2},1\big)&\Ib b\cr
}$$
but it is useful to keep it.

Define the ``propagator'' as the inverse of the quadratic
form in \Ia, represented as
$\propagarr  ={1\over N\alpha}\delta_{il}\delta_{jk}$ 
and the 4-vertex as the tensor
$\vertexarr=gN \delta_{qi}\delta_{jk} \delta_{l m}\delta_{np}$.
This four-vertex is to be considered as a rigid crossing, which cannot
be flipped 
and in which the cyclic order of the lines cannot be changed.
In both the propagator and the 4-vertex, the small arrows distinguish
the row and column indices of the matrices, while the wide one
distinguishes $M$ from $M^\dagger$. 

The prescriptions to compute the $n$-th order of the $g$-expansion of
$F$, known as Feynman rules, are as follows: draw $n$ four-vertices, 
 then draw all the topologically
distinct connected graphs obtained by joining  by propagators 
the double lines emerging from these $n$  4-vertices, while respecting the 
orientations, and  sum over the matrix indices $i,j,\ldots=1,\ldots,N$. 
 Each graph then comes with a weight $g^n N^\# /\alpha^{2n}$, where 
the power of $N$ will be computed below, and 
a ``symmetry factor'', 
which is the inverse of the order of the group of
permutations of the lines and vertices which leave the structure 
of the graph unchanged (see below an alternative characterization
of this factor). 

When drawing these Feynman diagrams on a plane, 
one usually  encounters topological obstructions which force one to
introduce additional crossings (over- or under-, it is immaterial).
Alternatively, these diagrams may be drawn on a higher genus Riemann 
surface as can be seen as follows. The identification of matrix 
indices by the Kronecker delta's of the propagators and 4-vertices 
leaves us with a  number $\#F$ of index loops. By pasting a domain
homeomorphic to a disk to each such loop, we build a 
discretized Riemann surface with $n$ edges, $2n$ edges and $\#F$ faces.

Thus Feynman diagrams for $F(g,\Ga)$ may be regarded as discretized 
orientable Riemann surfaces $\Sigma$.
Their faces are oriented by the small arrows carried by the propagators. 
In addition their edges carry the orientation of the big arrows. 
According to the argument of 't Hooft \tH\ 
and following the rules above, 
if a diagram has $\# V= n$ vertices, hence $\# E=2n$ edges (propagators), 
and $\# F$ faces,
 it carries a power of $N$ equal to $n-2n +\# F=\chi_E(\Sigma)=2-2h$, 
the Euler-Poincar\'e characterics expressed in terms of the genus $h$. 

To summarize, we have obtained a topological 
expansion (which is an asymptotic expansion in $1/N^2$) 

\eqn\Ftopo{ F(g,\Ga)=\sum_{h=0}^\infty {1\over N^{2h}} F^{(h)}(g,\Ga)}
where $ F^{(h)}$ is the sum over Feynman diagrams of genus $h$
weigthed as explained above. 
Now that this property has been established, we abandon 
the double line notation and return to more conventional notations
for Feynman diagrams: we erase the small arrows of matrix indices but 
retain the big ones that encode the distinction between $M$ and $M^\dagger$.

We are now ready to build a 
dictionary with virtual links: due to the ``contraction'' of 
$M$ and $M^\dagger$ through the propagators, 
Feynman diagrams of the type just discussed are naturally
endowed with the properties of alternating virtual link diagrams. 
Thus $4$-vertices 
are in one-to-one correspondence with over/under-crossings 
\vertex \ $\Leftrightarrow$ \vertexalt, 
while the virtual crossings admit the alternative representation:
\vcralt. (Beware! a virtual crossing is depicted in the graph 
theoretic way as an under- or over-crossing in the Feynman diagram
representation; as mentioned above, it is immaterial to draw it either 
way.) Note that in this representation, it is quite natural
that these virtual crossings can be freely moved around, 
thus enforcing the virtual Reidemeister moves. 
Note also that this correspondence gives an operative way to
compute the genus on which to draw a given virtual link diagram,
which may not have been obvious in their original presentation
(whether or not this is the {\it minimal}\/ genus on which one can draw the link itself
is a subtle matter due to the existence of the ``real'' Reidemeister moves, and it will
be discussed again in section 4).

From the relations $n=\# F+2h-2$ and $\# F\ge 1$, we might have expected
contributions of
genus $h$ to occur first at order $n=2h-1$, for diagrams with a single
face. This is indeed what happens with diagrams related to hermitian 
matrix integrals. Here, however, the orientation of edges by the big
arrows induces an additional constraint: adjacent faces have 
opposite orientations with respect to these arrows.  This forbids the
possibility to have $\# F=1$ and as a result, genus $h>0$ occurs first at
order 
\eqn\nmin{ n_{{\rm min}}(h)= 2h\ ,}
i.e. $F^{(h)}(g,\Ga)$ starts at order $g^{2h}$, or said otherwise, 
there is no virtual link of genus $h$ with less than $2h$ (real)
crossings. One checks that the bound is saturated by computing the 
coefficient, 
\eqn\Fmin{F_{2h}^{(h)} ={(4h-1)!!\over 4 h(2h+1)}\ .}
and it is also possible to determine without much further effort
the next term, i.e. the coefficient of $g^{2h+1}$
\eqn\Fnlo{F_{2h+1}^{(h)} =  {(4h+1)!!\over (2h+1)^2}
\sum_{s=0}^{2h} {(-1)^{\lfloor{s+1\over 2} \rfloor}
\over {4h+1\choose s}} {2h\choose
\lfloor{s\over 2} \rfloor }\sum_{p=s+1}^{4h+1-s}{1\over p}\ ,}
see Appendix A for details.

\subsec{Correlation functions}
We are also interested in the  ``$2p$-point functions'' $G_{2p}(g,\alpha)
:=\bra {1\over N}\tr (MM^\dagger)^p\ket$, in 
particular 
\eqn\Gfour{ G_4(g,\Ga)=2 {\partial F(g,\Ga)\over\partial g}}
and
\eqn\Gtwo{ G_2(g,\Ga) =
{1\over\alpha}-{\partial\over\partial\alpha}F(g,\alpha)=
{1\over\alpha}(1+g G_4)\ ,}
where use has been made of the homogeneity property \Ib{}. 
This same property implies that 
\eqn\homoG
{\eqalign{G_{2p}(g,\Ga)&={1\over \Ga^p}G_{2p}\({g\over\Ga^2}\)\cr
G_{2p}(g):&= G_{2p}(g,1)}}
These functions too admit a graphical representation, with similar 
Feynman rules and Feynman graphs with $2p$ external lines. These graphs 
are natural candidates for $2p$-tangle diagrams. For the 
four-point function $G_4$, we adopt the following 
convention of orientation: external lines may be extended 
to a circle surrounding the diagram, and the four lines are drawn
in the NW, NE, SE and SW directions, with the 
outcoming arrow (i.e. first crossing is over-) 
on NW and SE external lines. 

\subsec{Nugatory crossings and non-prime diagrams}
The two-point function $G_2$ is 
in particular useful  to dispose of all composite links and tangles.
Irrelevant ``nugatory'' crossings and non prime diagrams
appear as graphs  with a subgraph 
which may be disconnected by cutting transversely two distinct edges.  
Such a subgraph is called  a ``self-energy'' by physicists.
To remove all nugatory crossings and non prime configurations,  
i.e. to retain only diagrams with no self-energy, 
it suffices to choose  $\alpha=\alpha(g)$ so as to make 
$G_2(g,\alpha(g))=1$ or equivalently in view of \homoG, 
$G_2(g/\alpha(g)^2)=\alpha$, and to plug it into $F(g,\alpha)$ 
 or $G_{2p}(g,\Ga)$. Then the ``connected four-point function with 
no self-energy'' 
defined as 
\eqn\Gamdef{\Gamma(g)= G_4(g,\alpha(g))- 2}
 is easily seen to satisfy
\eqn\alphaGam{g \Gamma(g)= \alpha(g) -1 -2g } 
as a consequence of \Gtwo. (The appearance of  $-2$ in \Gamdef\  
is due to the
subtraction of disconnected contributions to the 4-tangle by two parallel 
non-intersecting strands.)

\subsec{Flypes}
\noindent
We now want to 
argue that dividing only by the {\it planar} flypes suffices
to get the equivalence classes of virtual alternating tangles
and links.
This claim is based partially on our intuition that 
other types of moves, such as flypes of higher genus, are
not permitted by the structure of the thickened 
surface, and partially on the study of low order virtual
links and tangles. But our best evidence comes from the
analysis, explained in section 4,
of several classes of invariants applied to links and tangles up
to order 6 (six real crossings), which indicates that the remaining
objects are indeed topologically inequivalent. Still
this remains an assumption\dots

For classical (genus 0) tangles, Sundberg and \Th\  have shown 
how to construct the generating function of flype-equivalence
classes of tangles $\tilde\Gamma^{(0)}(g)$ 
from the planar generating function 
$\Gamma^{(0)}(g)\equiv\lim_{N\to\infty}\Gamma(g)$ \STh. The 
operations leading from $\Gamma^{(0)}(g)$ to  $\widetilde\Gamma^{(0)}(g)$
have subsequently been shown by one of us \PZJ\ to be simply   
expressible in terms of a ``coupling constant renormalization''
in the language of physicists, i.e.\ of a redefinition of
the expansion variable, determined in a self-consistent way.
Let $g_0(g)$ be the solution of 
\eqn\gzero{
 g_0 = g\(-1 + {2\over (1 - g)(1 + \Gamma^{(0)}(g_0))}\) \ .}
Then we recover the result of Sundberg and \Th\ by
writing that $\widetilde\Gamma^{(0)}(g)=\Gamma^{(0)}(g_0(g))$.
According to our conjecture, we must generalize this to all genera.
If $\widetilde\Gamma(g)=\sum_{h=0}^\infty N^{-2h} \widetilde\Gamma^{(h)}(g)$
is the full generating function of virtual prime alternating tangles, then
we are led to
\eqn\miracle{
\widetilde\Gamma(g)=\Gamma(g_0(g))\ .
}


\newsec{Sum over all genera}
\noindent 
For $N=1$ the integral over a single real variable is readily 
computed. One finds for $Z$ the asymptotic expansion
\eqn\Zpert{Z(g,\alpha=1)\big|_{N=1}=\sum_{0}^\infty \({g\over 2}\)^n 
{(2n)!\over n!}\ .}
Its logarithm
\eqn\Fallgen{\eqalign{F(g,\alpha=1)\big|_{N=1}&=g + \frac{5}{2}\,g^2 + \frac{37}{3}\,g^3 
+ \frac{353}{4}\,g^4 + \frac{4081}{5}\,g^5  +
 \frac{55205\,g^6}{6}\,g^6  
 + \frac{854197}{7}\,g^7 \cr& \qquad + \frac{14876033}{8}\,g^8 + 
  \frac{288018721}{9}\,g^9 + \frac{1227782785}{2}\,g^{10}
+\frac{142882295557}{11}\, g^{11}
+O(g^{12})}}
is the sum of $F^{(h)}$ for all  genera $h$ 
and will provide a sum rule over the contributions of the different 
genera to be discussed in the next section.
By differentiation one gets the 4-point function
\eqnn\Gfourag
$$\eqalignno{G_4(g)&=2 + 10\ g + 74\ g^2 + 706\ g^3 + 8162\ g^4 
+ 110410\ g^5 +     1708394\ g^6 & \Gfourag
\cr &\qquad+ 29752066\ g^7 + 576037442\ g^8 + 12277827850\
g^9 +285764591114\ g^{10}+O(g^{11})}$$
and after determination of $\Ga(g)$ as explained in sect.~1.3, one
gets the connected four-point function with no self-energy 
\eqnn\Gamag
$$\eqalignno{\Gamma(g)=&2\ g + 10\ g^2 + 82\ g^3 + 898\ g^4 +
12018\ g^5 + 187626\ g^6   & \Gamag\cr&+
    3323682\ g^7 + 65607682\ g^8 + 1424967394\ g^9 +33736908874\
g^{10}+O(g^{11})\ .}$$

Expanding Eq.~\gzero\ to the required order, we find
\eqn\gzero{
g_0=g - 2\,g^3 - 4\,g^4 - 10\,g^5 - 30\,g^6 - 108\,g^7 -436 g^8-1890 
g^9 -8588 g^{10} +O(g^{11})}
and therefore
\eqnn\Gamtag
$$\eqalignno{\widetilde\Gamma(g)=&
2\ g + 10\ g^2 + 78\ g^3 + 850\ g^4 +
11426\ g^5 + 
179238\ g^6 & \Gamtag\cr  &+ 3187002\ g^7  +
63095526\ g^8  + 1373767142\ g^9  + 3259401885\ g^{10}
+O(g^{11})\ .}$$

The three expansions $G_4(g)$, $\Gamma(g)$ and $\widetilde\Gamma(g)$
have the same asymptotic behavior up to a multiplicative constant, 
with their $n$-th order of the form ${\rm const.\ } 2^n \sqrt{n+1} \, (n+1)!$

\newsec{Genus 0, 1, 2 and 3}
\noindent
We now return to the matrix integral \Ia\ and its $1/N^2$  expansion. 
Let  $a^2(g)$ be the solution of
\eqn\adeu{ a^2=1+ 3 {g} (a^2)^2}
with $a^2(g)=1+O(g)$. (Its interpretation is that it characterizes the 
support of the limiting distribution of eigenvalues of $M M^\dagger$.)
Then one finds 
\eqnn\Fzer
$$\eqalignno{F^{(0)}(g):=F^{(0)}(g,1)&=\log a^2-{1\over
12}(a^2-1)(9-a^2)\cr
&=2\sum_{n=1}^\infty  (3g)^n  {(2n-1)!!\over n!(n+2)!}\cr}$$
(As a side-remark, we recall that this is 
twice the result for the Hermitian matrix integral as in \Ia\ but with 
$\tr \({\Ga\over 2} M^2 -{g\over 4} M^4\)$ in the exponential).
Explicitly, one gets the expansion
\eqn\Fzero{\eqalign{F^{(0)}(g)&=
g + \frac{9}{4}\,g^2 + 9\,g^3 + \frac{189}{4}\,g^4 + \frac{1458}{5}\,g^5 + 
  \frac{8019}{4}\,g^6 + \frac{104247}{7}\,g^7\cr &\qquad 
+ \frac{938223}{8}\,g^8 +   966654\,g^9 + \frac{82648917}{10}\,g^{10}
+\frac{801058734}{11}\,g^{11}
+O(g^{12})\ .}}
For genus $1$, Morris gives \Mo
\eqnn\Fun
$$\eqalignno{F^{(1)}(g)&=-{1\over 24} \log {(2-a^2)(2+a^2)^3\over 27}\cr
& = {1\over 24}\sum_{n=0}^\infty {(3g)^{n+1}\over n+1}
\sum_{p=0}^n {(2n+2)!\over (n-p)!(n+2+p)!}\(1-(-3)^{-p}\)\cr
&=\frac{1}{4}g^2 + \frac{10}{3}\,g^3 + \frac{307}{8}\,g^4 + 428\,g^5 + 
  \frac{28457}{6}\,g^6 + 52612\,g^7 + \frac{9370183}{16}\,g^8
\cr & \qquad \qquad + 
  \frac{58911256}{9}\,g^9 + \frac{734641583}{10}\,g^{10}
+ 827733428\,g^{11} +O(g^{12}) \ .
\cr}$$

For higher genus, the expressions are more and more complicated.
Akemann and Adamietz \refs{\Ake,\Ada} have found that in terms of 
$I_1=1-6 g a^2$ and $M_0=1 -2 g a^2$ 

\eqnn\Fdeu
$$\eqalignno{F^{(2)}(g) &=  
{21 a^2 g^3 \over 40 I_1^5} - { 69 g^2\over 640 I_1^4}
+ {53 g \over 2560 a^2 I_1^3} + {g \over 256  a^2 I_1^2 M_0} \cr
&\qquad -{3 g\over 512 a^2 M_0^3} -{1\over 512 a^4 I_1 M_0}
-{3\over 1024 a^4 M_0^2} -{53\over 15360 a^4 I_1^2 } \cr} $$
%
whence
\eqn\Fdeux{\eqalign{F^{(2)}(g)&=
\frac{21}{8}\,g^4 + \frac{483}{5}\,g^5 + \frac{4659}{2}\,g^6 +
46434\,g^7 \cr & \qquad + 
  \frac{6635991}{8}\,g^8 + 13798410\,g^9 + \frac{1091610282}{5}\,g^{10}
+3328687092\,g^{11}+O(g^{12})}}
and the expression of $F^{(3)}(g)$ is too cumbersome to be given here
but leads to the expansion
\eqn\Ftroi
{F^{(3)}(g)=
\frac{495}{4}\,g^6 + \frac{56628}{7}\,g^7 + \frac{2504115}{8}\,g^8 + 9322668\,g^9 + 
  \frac{472138479}{2}\,g^{10}+ 5345163216\,g^{11}
+O(g^{12}) \ .}
\ommit{x = 4 a2; (* notations d'Adamietz *)F3A = 
  Series[2205 x^2 I2^6/(256 I1^10) + 405 x I2^5/(256 I1^9) + 
      765 I2^4/(512 I1^8) + 63 I2^4/(256 M0 I1^7) + 8985 I2^3/(7168 x I1^7) + 
      27 I2^3/(128 x M0 I1^6) + 2995 I2^2/(3584 x^2 I1^6) + 
      55 I2^2/(1024 x^2 M0^2 I1^4) + 201 I2^2/(1024 x^2 M0 I1^5) + 
      599 I2/(1344 x^3 I1^5) + 9I2/(512  x^3 M0^3 I1^2) + 
      85 I2/(1024 x^3 M0^2 I1^3) + 19 I2/(128 x^3 M0 I1^4) + 
      63 M1^2/(1024 x^2 M0^6) - 9 M1 I2 /(1024 x^2 M0^4 I1^2) - 
      117 M1/(1024 x^3 M0^5) - 9 M1/(512 x^3 M0^4 I1) + 599/(4032 x^4 I1^4) + 
      19/(256 x^4 M0 I1^3) + 85 /(1024 x^4 M0^2 I1^2) + 
      45/(512 x^4 M0^3 I1) + 117/(1024 x^4 M0^4), {g, 0, nt} ]}

We check that these $F^{(h)}$ start at an order in $g$ consistent 
with \nmin. Moreover the sum of these four first contributions
differ from the sum over all genera \Fallgen\ by terms of order 
$g^8$ as it should. Using the additional information of \Fmin\ and \Fnlo, 
one may extract the first terms of $F^{(4)}$ and $F^{(5)}$
\eqn\Ffourf{
\eqalign{F^{(4)}&=
\frac{225225}{16}\,g^8 + 1368653\,g^9 + \frac{1495900107}{20}\,g^{10}
+3023618067\,g^{11}  +O(g^{12})\cr
F^{(5)}&=\frac{11904165}{4}\, g^{10}
+ \frac{4304016990}{11}\,g^{11}
 +O(g^{12})\ .\cr
}}
By differentiating with respect to $g$, one gets $G^{(h)}_2(g,1)$ and
$G^{(h)}_4(g,1)$ according to \Gtwo\ and \Gfour.
One then determines the double expansion in powers of $g$ and $1/N^2$ 
of $\Ga(g)$ so as to remove the self-energies, as explained at the end of 
sect. 1. We don't display the corresponding expansion of $\Ga(g)$
as it may be recovered from eqn \alphaGam\ and the expressions
of $\Gamma^{(h)}$ below. Using the additional data of \Ffourf\
we may provide the 
$g$-expansion up to order $g^{10}$ of $\Gamma^{(h)} (g)$ for $h=0,\ldots,5$. 
\eqnn\Gamm
$$\eqalignno{\scriptstyle{
\Gamma_{}^{(0)}(g)
}& \scriptstyle{
= g + 2\,g^2 + 6\,g^3 + 22\,g^4 + 91\,g^5 + 408\,g^6 + 
1938\,g^7 + 9614\,g^8 +   49335\,g^9 +260130\,g^{10}+O(g^{11})
}\cr \scriptstyle{
\Gamma_{}^{(1)}(g) 
}& \scriptstyle{ 
= g + 8\,g^2 + 59\,g^3 + 420\,g^4 + 2940\,g^5 + 20384\,g^6 
+ 140479\,g^7 +   964184\,g^8 + 6598481\,g^9
+45059872\,g^{10}
+O(g^{11}) 
}\cr \scriptstyle{
\Gamma_{}^{(2)}(g) 
}& \scriptstyle{
=17\,g^3 + 456\,g^4 + 7728\,g^5 + 104762\,g^6 + 1240518\,g^7 + 13406796\,g^8 + 
  135637190\,g^9 + 1305368592\,g^{10} +O(g^{11})
}\cr \scriptstyle{
\Gamma_{}^{(3)}(g) 
}& \scriptstyle{
= 1259\, g^5 + 62072\, g^6 + 1740158\, g^7 + 36316872\, g^8 + 
    627368680\, g^9 + 9484251920\,g^{10}  +O(g^{11})
} & \Gamm \cr \scriptstyle{
\Gamma_{}^{(4)}(g) 
}& \scriptstyle{
=200589\ g^7 + 14910216\ g^8 + 600547192\ g^9
+ 17347802824\,g^{10}+O(g^{11})
}\cr \scriptstyle{
\Gamma_{}^{(5)}(g) 
}& \scriptstyle{
=54766516\ g^9+ 5554165536\,g^{10}+O(g^{11})
}}$$
are the generating functions of the numbers of connected graphs of
genus 0 to 3 with no self-energy; they 
are not yet the generating functions of the number of tangle diagrams, 
due to the flype equivalence and possible other redundancies. 
One may integrate these expressions according to \Gfour\ to obtain the 
corresponding generating functions $F^{(h)}_{}(g)$ of 
virtual alternating link diagrams with no self-energy. 

In figures \IPItwo$\,$-\IPIfourc{} we depict the corresponding diagrams of $F_{}^{(h)}$,
$h=0,1,2$ up to order 4: the corresponding diagrams of $\Gamma_{}$
are obtained by removing in all possible non equivalent ways one
vertex, thus opening the link diagram into a tangle. In these figures, 
we list in parallel the two notations of Feynman diagrams and of
links. In the latter, colors have been introduced only to distinguish 
the different connected components.
Each link of order $n$ ($n$ crossings) comes with an 
integer, whose inverse gives its weight in $F_{}^{(h)}$.
Alternatively, the number of distinct contributions
that this link gives to $\Gamma_{}^{(h)}$ after removal of one
vertex equals $2n$ divided by this integer.

\fig{The genus 0 and 1 2-crossing alternating virtual link diagrams 
in the two representations, the Feynman diagrams on the left, the virtual
diagrams on the right: for
each, the inverse of the weight in $F$ 
is indicated}{\epsfbox{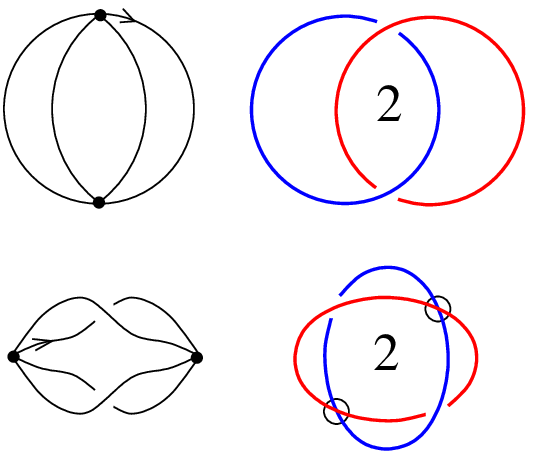}}{4cm}\figlabel\IPItwo

\fig{same for order 3, genus 0 and 1}{\epsfbox{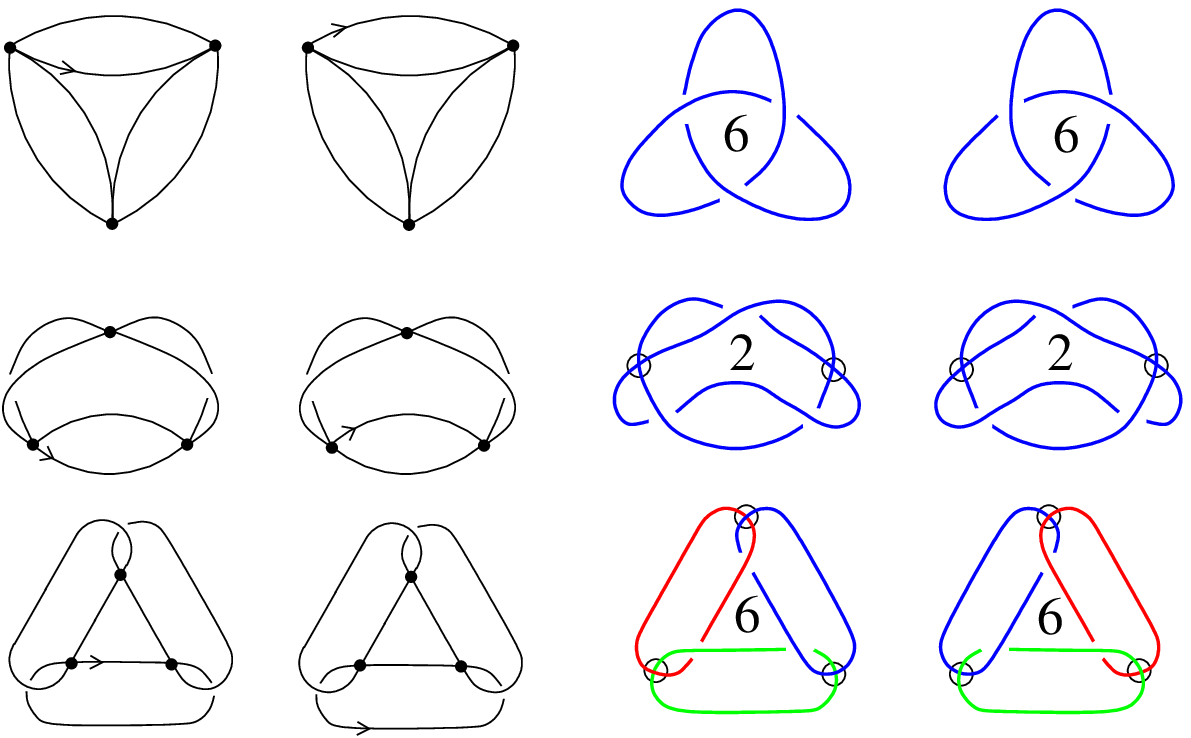}}{8cm}
\figlabel\IPIthree

\fig{same for order 4, genus 0 }
{\epsfbox{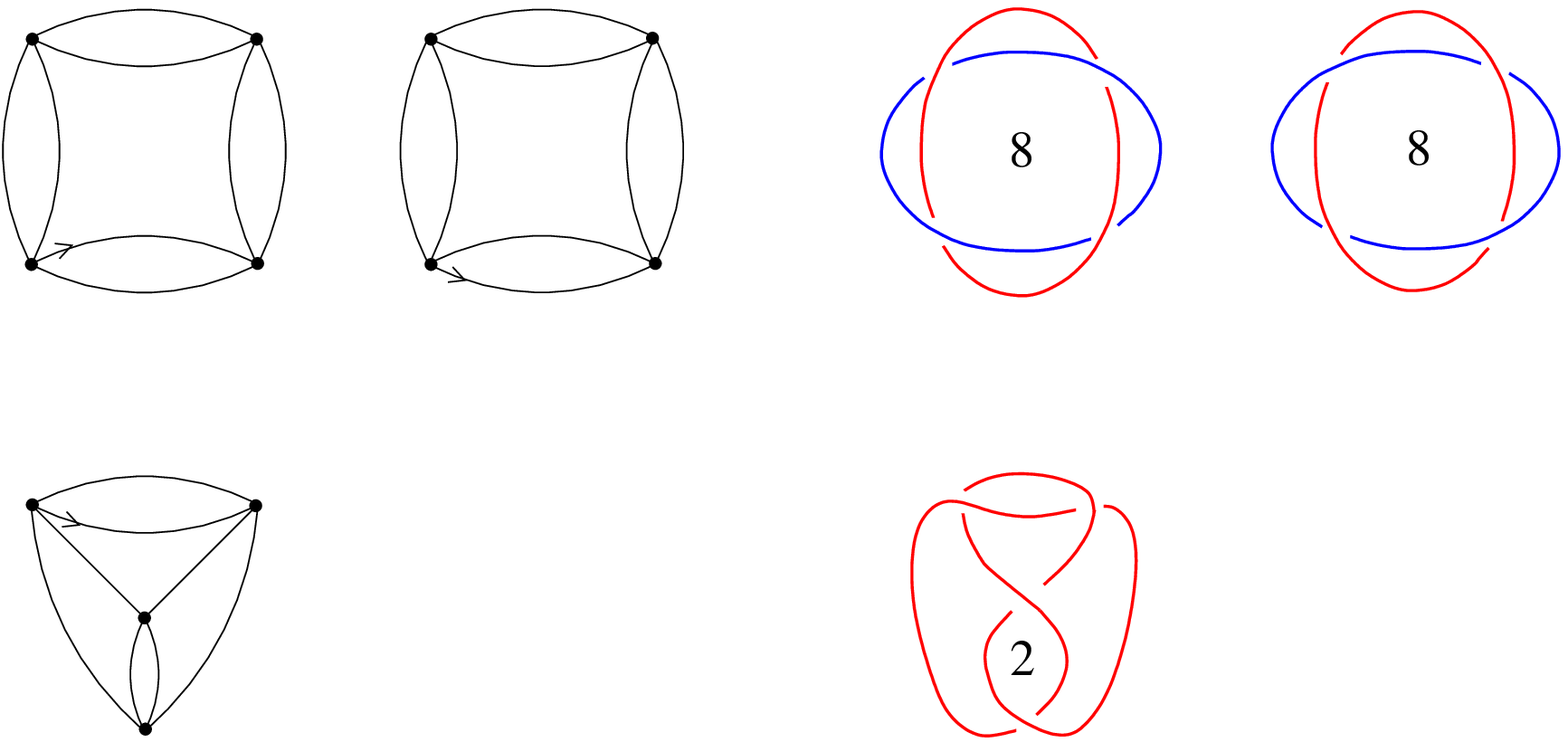}}{10cm}\figlabel\IPIfour{}

\fig{{\it ibid} for genus 1}{\epsfbox{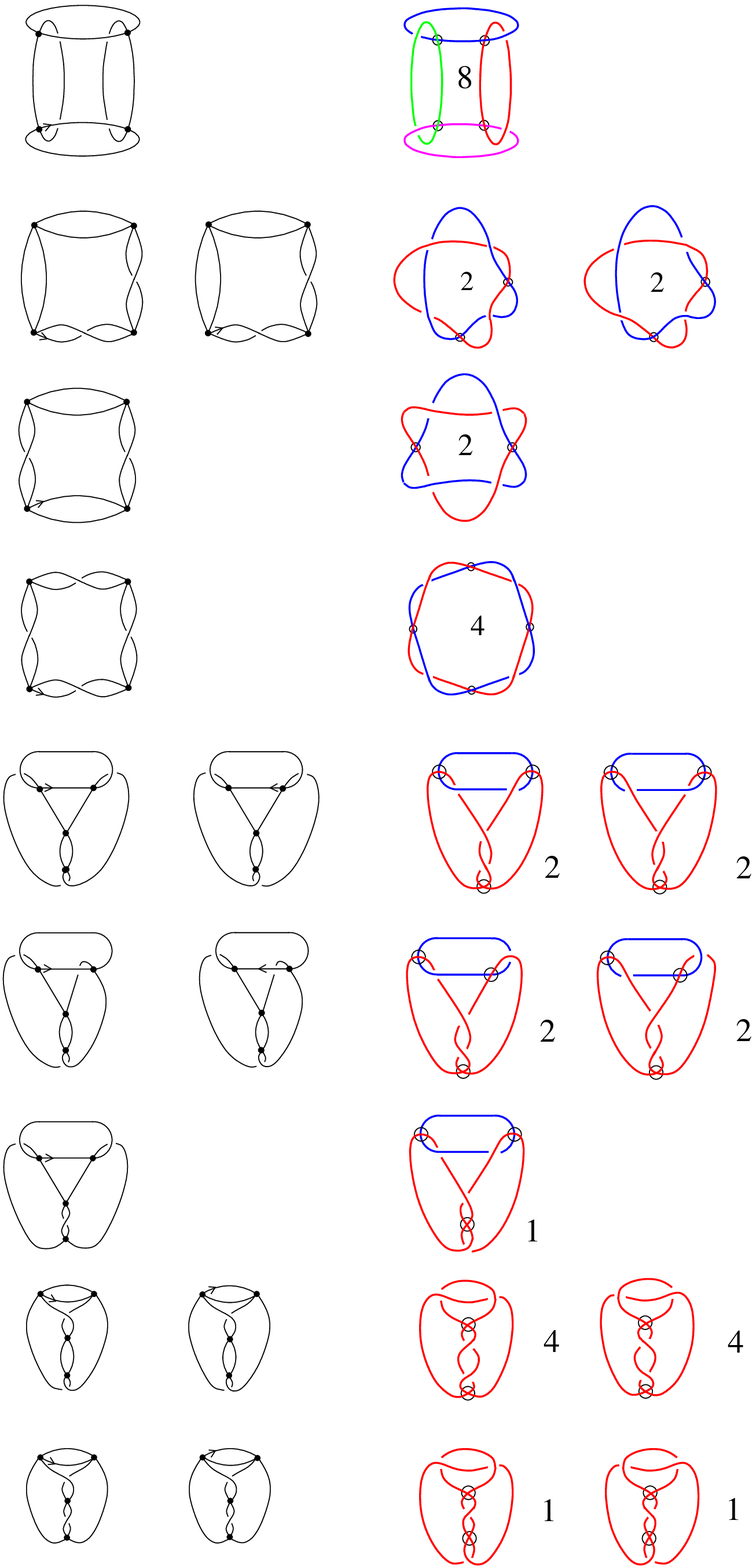}}{10cm}\figlabel\IPIfoura{}

\fig{{\it ibid} for genus 2}{\epsfbox{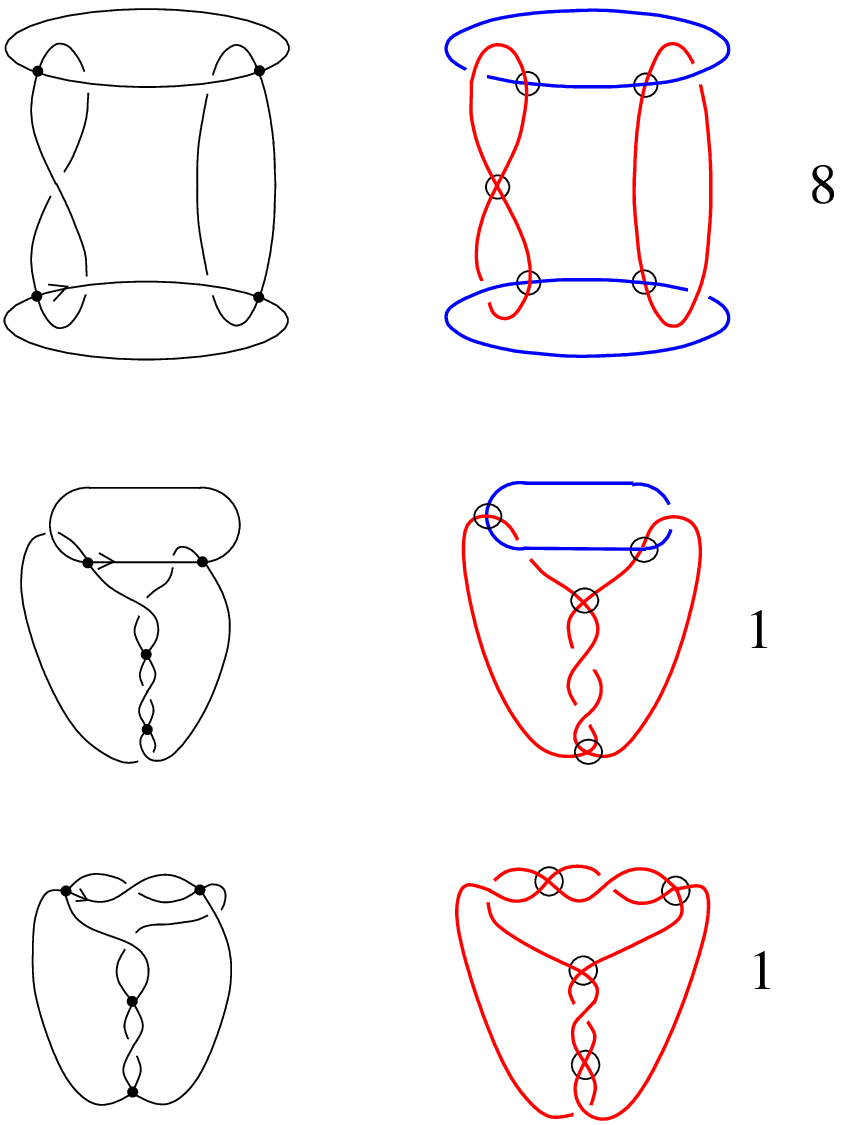}}{6cm}\figlabel\IPIfourc{}

\fig{First occurences of flype equivalence in tangles with 3
crossings}{\epsfbox{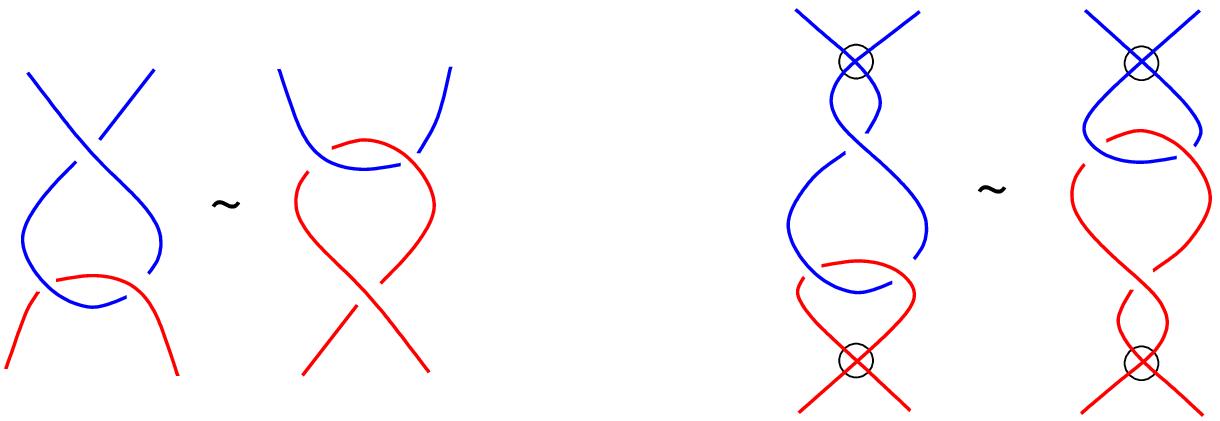}}{8cm}\figlabel\firstflyp

The first flypes occur at order 3 for genus 0 or genus 1 in $\Gamma$, 
see fig. \firstflyp. 
The generating functions of flype equivalence
classes of classical links is
$$\widetilde{\Gamma}^{(0)}(g)=
\Gamma_{}^{(0)}(g_0)=g + 2\,g^2 + 4\,g^3 + 10\,g^4 + 29\,g^5 + 98\,g^6 
+ 372\,g^7  + 1538 g^8 + 6755 g^9 + 30996\,g^{10}+ O(g^{11})$$
(which is the result of \STh).  Then, according to our assumption,
\eqnn\Gamtild
$$\eqalignno{\widetilde{\Gamma}^{(1)}(g)&=
\Gamma_{}^{(1)}(g_0)=g + 8\,g^2 + 57\,g^3 + 384\,g^4 + 2512\,g^5 \cr
& \qquad\qquad + 16158\,g^6 + 102837\,g^7 + 649862\,g^8 + 
  4086137\,g^9+ 25597900\,g^{10} +O(g^{11}) \cr
\widetilde{\Gamma}^{(2)}(g)&=\Gamma_{}^{(2)}(g_0)=
17\,g^3 + 456\,g^4 + 7626\,g^5 + 100910\,g^6 + 1155636\,g^7 +
11987082\,g^8 \cr
&  \qquad\qquad +   115664638\,g^9+ 1056131412\,g^{10} +O(g^{11})
\cr
\widetilde{\Gamma}^{(3)}(g)&=\Gamma_{}^{(3)}(g_0)=
1259\,g^5 + 62072\,g^6 + 1727568\,g^7 + 35546828\,g^8 + 601504150\,g^9
\cr 
 &  \qquad\qquad +  8854470134\,g^{10}
 +O(g^{11})   & \Gamtild
\cr
\widetilde{\Gamma}^{(4)}(g)&=\Gamma_{}^{(4)}(g_0)=
200589\ g^7 + 14910216\ g^8 + 597738946\ g^9+ 17103622876\,g^{10}
 +O(g^{11})
\cr
\widetilde{\Gamma}^{(5)}(g)&=\Gamma_{}^{(5)}(g_0)=
54766516\ g^9+ 5554165536\,g^{10} +O(g^{11})
\ }$$
are the generating functions of flype-equivalence classes of virtual 
tangles with up to 10 real crossings. For example, the reduction from 
59 to 57 of the number of genus 1 tangles
of order 3 is in accordance with the equivalence of fig. \firstflyp.

The large order behavior of the $g$-expansions of $F^{(h)}$ is
dominated by the leading singularity of $a^2(g)$, 
which occurs at $g=g_c=1/12$, $a^2(g_c)=2$. One finds 
\eqn\singbeh
{\eqalign{F^{(0)}(g,1)&\approx \(g_c-g\)^{5/2}\cr
F^{(1)}(g,1)&\approx \log\(g_c-g\)\ . } }
The leading singularity of the expression of  $F^{(2)}$ comes from
the term proportional to $I_1^{-5}$, as $I_1$ vanishes like
$(g_c-g)^{1/2}$. Similarly $F^{(3)}$ has a pole of order 10 in $I_1$.
This is typical of what is expected for generic genus
\DFGZJ\ 
\eqn\singul{ F^{(h)}(g)\approx  \(g_c-g\)^{5/2(1-h)}\ .}
Thus one expects
\eqn\largeord{f_n^{(h)}\approx  {1\over g_c^n}   n^{5/2(h-1)-1}\ , }
and correspondingly, for $\Gamma^{(h)}(g)= \sum_n \gamma_n^{(h)} g^n$
\eqn\Gammasym{\gamma_n^{(h)}\approx  {1\over g_c^n}   n^{5/2(h-1)}\
. }
The subsequent reductions that we perform to eliminate
the redundancies, change the value of $g_c$ (enlarging it
so as to increase the radius of convergence) but do not affect 
the value of the ``critical exponents'' $5/2,0, -5/2, \dots$
Thus, removing the self-energies has the effect that 
the closest singularity is now for $g/\alpha^2(g)=1/12$ 
which gives \ZJZun\ $g'_c=4/27$. 
Similarly, taking care of the flype equivalence
increases the radius of convergence of $\widetilde{\Gamma}(g)$
to the value $ g''_c=(-101+\sqrt{21001})/270 $, see \STh, 
but does not affect the general form
\Gammasym. 

Finally, note that the function $\widetilde{F}(g)$ obtained by
integrating $\widetilde{\Gamma}$ is not exactly the generating 
function of flype-equivalence classes of links (due to the
issue of symmetry factors) but should 
have the same asymptotic behavior: intuitively, this reflects the
fact that the number of link diagrams with a non trivial symmetry
factor is subleading and does not contribute to the asymptotic
behavior of the form \largeord\ (this fact has now been proved for
classical links \KJS).


\newsec{An algorithm to classify virtual alternating links}
\noindent
In this section we describe the encoding we used to represent
virtual alternating link diagrams,
and the subsequent algorithm that allowed us to generate prime
alternating virtual links. Due to
the factorial growth of their number, we only describe the result up
at order 4; but we have obtained data up to order 6 in order to check
our generalized flype conjecture.

\subsec{Alternating link diagrams and permutations}
\noindent Our encoding of virtual
alternating link diagrams is based on a well-known correspondence
between bicolored maps and permutations. 
An alternating link diagram
can be equivalently described as a (not necessarily planar) map whose
vertices have valence 4 and whose faces are bicolored, according to
the pattern of under/over crossings as one moves around the face,
see Fig.~\bico.
\fig{Bicoloration and labelling 
of an alternating link diagram.}{\epsfbox{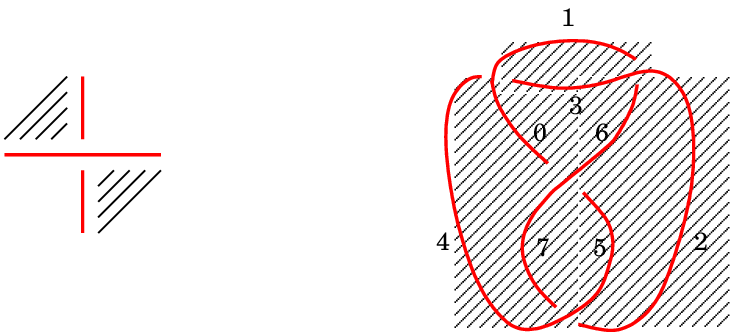}}{5cm}\figlabel\bico

Let us now {\it label}\/ the edges of the diagram (or of the map): 
the set of edge labels will
be called $E$, and its permutation group ${\cal S}(E)$. 
In the implementation,
$E$ is chosen to be
$E=\{0,1,\ldots,2n-1\}$. It is known that general face-bicolored maps
(i.e.\ duals of bipartite maps) are
in one-to-one correspondence to pairs of permutations $(\sigma,\tau)\in{\cal S}(E)$
according to the following recipe: the cycles of $\sigma$ (resp.\ $\tau$)
are the labels of the
edges in their cyclic order as one turns clockwise around white faces
(resp.\ counterclockwise around black faces).\foot{In terms of the
Feynman diagrams of the matrix model, $\sigma$ and $\tau$ correspond
to following either of the two lines of an edge in the direction
of its (big) arrow.}
Define additionally $\rho=\sigma^{-1} \tau$ and $\tilde{\rho}=\sigma\tau^{-1}$.
The cycles of $\rho$ (or of $\tilde{\rho}$) are easily seen to be in one-to-one
correspondence with vertices of the map.

Finally a relabelling of the map is a permutation $g\in{\cal S}(E)$ of the
labels acting by conjugation:
\eqn\permconj{
\sigma'=g\sigma g^{-1}\qquad \tau'=g\tau g^{-1}
}
An unlabelled map can therefore be described as a conjugacy class of pairs
of permutations.

Here we require various additional properties of the map, which must
be translated combinatorially into properties of the permutations:

1) First and foremost, all vertices must have valence 4. This implies
that $\rho$ (resp.\ $\tilde{\rho}$) only has
2-cycles, i.e.\ is a fixed point-free involution, exchanging edges
at overcrossings (resp.\ undercrossings).
Here we decide
to focus on $\rho$ rather than $\tilde{\rho}$. The situation at each
vertex is described on Fig.~\rule; the figure can be considered as the defining
rule to build $\sigma$ and $\tau$.
\fig{Configuration at a vertex. If $\beta=\rho(\alpha)$, consistency
implies that $\sigma(\alpha)=\tau(\beta)$ and $\tau(\alpha)=\sigma(\beta)$
 i.e.\ that
$\rho=\tau^{-1}\sigma=\sigma^{-1}\tau$ is an involution.}%
{\pic(924,1206)(1189,-1024){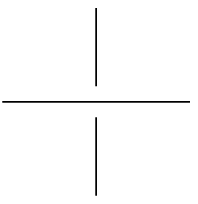}%
\put(1276,-586){$\alpha$}%
\put(1726,14){$\tau(\alpha)$}%
\put(2026,-586){$\rho(\alpha)$}%
\put(1726,-961){$\sigma(\alpha)$}%
}{1.5cm}\figlabel\rule

One can partially fix the freedom on the labels by noting that via conjugations,
one can reduce $\rho$ to a given form; for $E=\{0,1,\ldots,2n-1\}$, we choose
\eqn\permrho{
\rho(2\alpha)=2\alpha+1\qquad \rho(2\alpha+1)=2\alpha\qquad \alpha=0,\ldots,n-1
}

Once $\rho$ is fixed, the data of $\sigma$ alone suffices to describe
the alternating link diagram since $\tau=\sigma\rho$.
Furthermore, all relabellings must commute with $\rho$; they form a group
$G=\{ g\in {\cal S}(E) \mid g\rho=\rho g\}$ which is isomorphic to
${\cal S}_n \times {\Z_2}^n$.

For example, the labelled diagram of Fig.~\bico\ is such that
$\rho$ is of the form of Eq.~\permrho, and we find 
\eqn\permex{
\sigma=\pmatrix{0&1&2&3&4&5&6&7\cr 3&4&1&6&2&7&0&5}
\qquad\tau=\pmatrix{0&1&2&3&4&5&6&7\cr 4&3&6&1&7&2&5&0}
}
or in terms of cycles $\sigma=(0\ 3\ 6)(1\ 4\ 2)(5\ 7)$ and 
$\tau=(0\ 4\ 7)(2\ 5\ 6)(1\ 3)$.

2) We are interested in {\it connected}\/ maps. This amounts to requiring that
the action on $E$ of the group generated by $\sigma$ and $\tau$
be transitive.

3) We mostly focus on diagrams without self-energy.
In order to find self-energies 
(i.e.\ subdiagrams with 2 external legs), we look for 
pairs of edges $(\alpha,\beta)$ which belong to the same cycle of $\sigma$ 
and to the same cycle of $\tau$.  Cutting these two edges amounts to 
composing with the transposition $(\alpha \beta)$: 
 $\sigma'=\sigma\circ (\alpha\beta)$, $\tau'=\tau\circ (\alpha\beta)$. 
The diagram has no self-energy iff for all such pairs $(\alpha,\beta)$, 
the modified diagram corresponding to $(\sigma',\tau')$ is still connected 
(note that in the planar case the resulting 
diagram is necessarily disconnected, so that the existence of such a pair
is enough to discard the diagram). 
\fig{Potential self-energies and how to cut them out.}{%
\pic(3644,1292)(429,-797){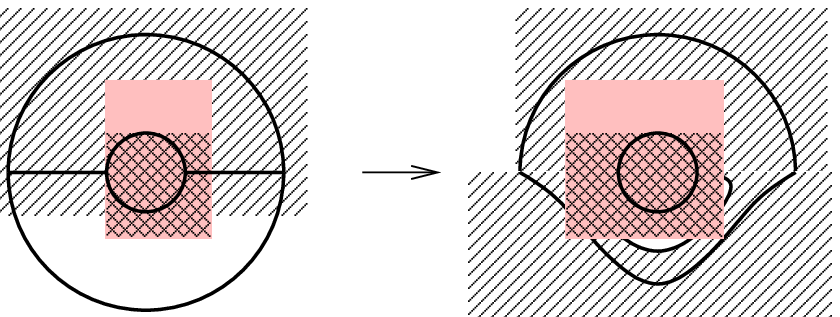}%
\put(631,-331){$\alpha$}%
\put(1401,-331){$\beta$}%
\put(2950,-541){$\alpha$}%
\put(3550,-331){$\beta$}%
}{6cm}

4) Finally, we want to consider classes of flype-equivalent diagrams. 
The flype acts on a diagram as follows:
consider four edges $(\alpha,\beta,\gamma,\delta)$ in the configuration 
depicted on Fig.~\permflyp a),
 that is $\sigma(\alpha)=\beta$, $\delta$ and $\gamma$ in the
same cycle of $\sigma$, $\alpha$ and $\delta$, $\beta$ and $\gamma$
in the same cycles of $\tau$.
Cut the tangle by composing $\sigma$ and $\tau$ with appropriate cycles,
and paste its legs together in the way described on Fig.~\permflyp b).
Proceed only if the resulting subdiagram is planar. 
If it is, then ``flip'' it by replacing $\sigma$ and $\tau$ with their 
inverses inside it, see Fig.~\permflyp c). 
Finally, reconnect the tangle to the rest of the diagram, see Fig.~\permflyp d).
A similar operation can be performed by exchanging black and white colors,
i.e.\ $\sigma$ and $\tau$ in the construction above.
Together these two types of moves reproduce all possible flypes.

\fig{Performing a flype via permutations.}{%
\pic(8326,1471)(609,-1100){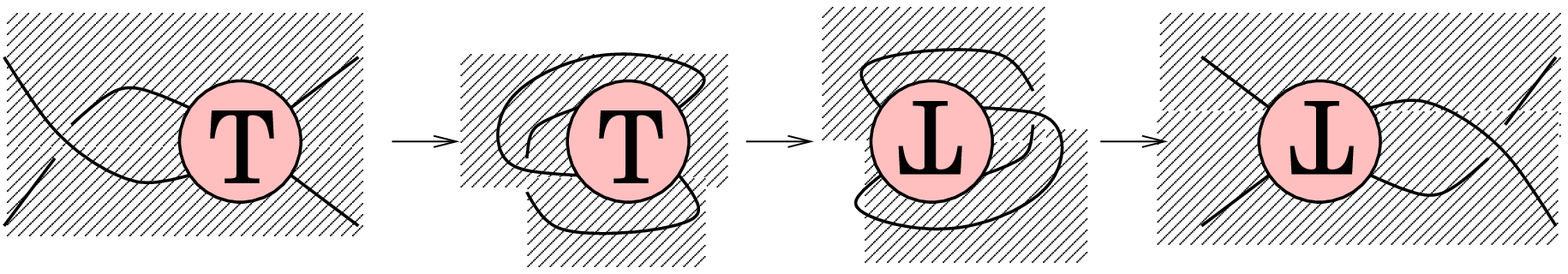}%
\put(1261,-601){$\nu$}%
\put(1261,119){$\mu$}%
\put(2521,-601){$\gamma$}%
\put(2521, 29){$\delta$}%
\put(411,119){$\alpha$}%
\put(411,-691){$\beta$}%
\put(4411,119){$\delta$}%
\put(4411,-781){$\gamma$}%
\put(3535,-565){$\nu$}%
\put(3433,-73){$\mu$}%
\put(5203,281){$\nu$}%
\put(5209,-835){$\mu$}%
\put(5923,-31){$\gamma$}%
\put(5971,-517){$\delta$}%
\put(6931, 41){$\alpha$}%
\put(6973,-589){$\beta$}%
\put(8929,-577){$\gamma$}%
\put(8935, 53){$\delta$}%
\put(8089, 59){$\mu$}%
\put(8113,-685){$\nu$}%
\put(4600,-100){flip}%
\put(3871,-1051){b)}%
\put(1531,-1051){a)}%
\put(5581,-1051){c)}%
\put(7831,-1051){d)}%
}{12cm}\figlabel\permflyp

\subsec{Calculation of link invariants}
\noindent 
There are various quantities one may want to compute once a permutation
$\sigma$ has been produced. They can be of many different types: first,
they may be true invariants of virtual links, or
they may be flype-invariant
and therefore conjectured invariants of reduced alternating virtual
link diagrams,
or they may be not invariant at all (but still interesting to
compute). Secondly, they may be invariants of {\it unoriented}\/ or {\it oriented}\/
links. In all that preceded we have only dealt with unoriented objects; however
many useful invariants depend on orientation and it is therefore
necessary to consider every choice of orientation ($2^c$ where $c$ is the
number of connected components) of an unoriented object.
We now list the quantities we have been able to compute,
and how to extract them from the permutation $\sigma$:

(i) The number of crossings $n$ is of course not left invariant
by Reidemeister moves,
but it is preserved by flypes. For reduced alternating
diagrams of virtual links it is conjectured to be the minimal number
of crossings.

(ii) The genus $h$ of the underlying surface: it is {\it not}\/ left invariant
by general Reidemeister moves, as Fig. \notorus\ shows 
(intuitively, after a Reidemeister move
a handle may become empty so that it must be removed),
however it is preserved by flypes, and once again, 
conjectured to be the minimal genus for virtual alternating links.
It is given by the Euler--Poincar\'e formula:
$\chi_E=2-2h=\# V-\# E+\# F$, where $\# V=n$ is the number
of vertices, $\# E=2n$ is the number of edges, $\#F$ is the number of 
faces.
If $\#\sigma$ is the number of cycles of $\sigma$ i.e.\ of white
faces, and similarly for $\tau$, we have $ \# F= \#\sigma +\#\tau$ and 
then we conclude that
\eqn\permgen{
h=1-{1\over2}(\#\sigma+\#\tau-n)
}
\fig{A genus 1 diagram  which turns out to be a trivial knot} 
{\epsfbox{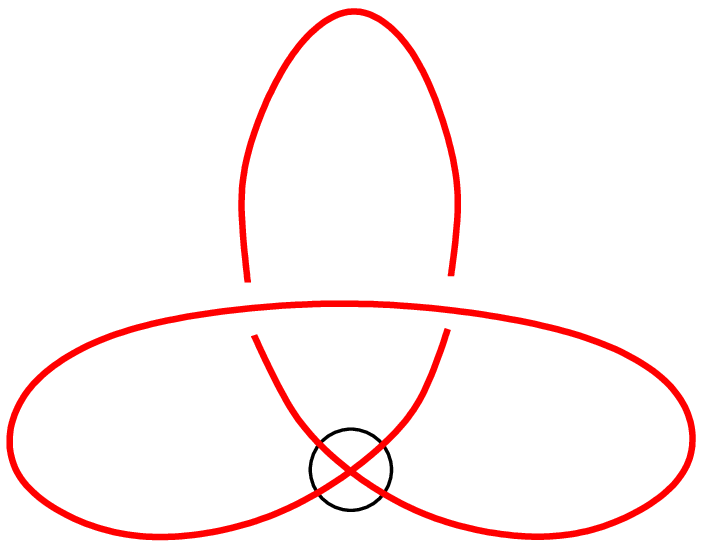}}{3cm}\figlabel\notorus

(iii) The number of connected components $c$ is of course an 
invariant of unoriented
virtual links. Moving along each connected component on the diagram can
be achieved by acting with $\rho$ and $\tilde{\rho}$ alternatingly;
it is easy to check that this implies
\eqn\permcc{
c={1\over2} \#(\rho\tilde{\rho})={1\over 2}\#(\sigma^{-2}\tau^2)
}

(iv) The order of symmetry of the diagram: this is not an invariant at all.
It is simply the order of the group of permutations $H$ that commute
with both $\sigma$ and $\tau$, that is
\eqn\permsubg{
H=\{g\in G \mid g\sigma=\sigma g\} \ .}
This order is a divisor of $2n$
(this results from the fact that for tangles -- see below --
this group is trivial).

At low orders one can easily find pairs 
of flype-equivalent alternating reduced diagrams with
distinct symmetry factors; this observation is important because it prevents us from
computing the generating function of the number of prime alternating links 
in the same way as for tangles. 

(v) The set of linking numbers: for an oriented diagram,
define the sign $\epsilon_v$ of a vertex $v$
according to Fig.~\ori, and
\eqn\permln{
\ell_{ij}=\sum_{v\in V_{ij}} \epsilon_v\qquad 1\le i, j\le c
}
$V_{ij}$ being the set of vertices where components labelled $i$ and $j$ meet.
\fig{Sign of a vertex.}{\epsfbox{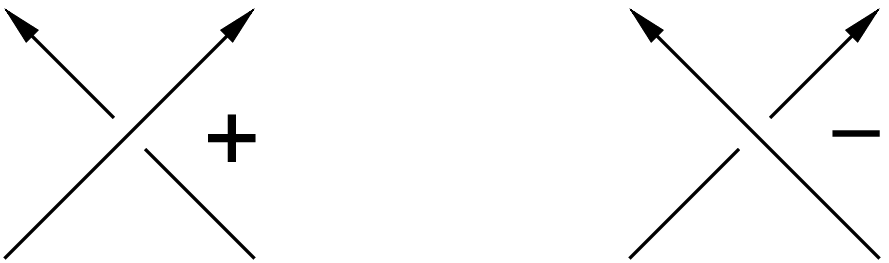}}{4cm}\figlabel\ori%

The off-diagonal elements $\ell_{ij}$, $i<j$ are twice the usual linking
numbers between components $i$ and $j$ (but they are not necessarily even
for virtual diagrams). They are invariants of oriented links (up to permutation
of the labels of the connected components). Their absolute value does not depend on orientation.

(vi) The determinant $d$ is an invariant of unoriented links: it is
a specialization of the usual Alexander polynomial (see below).

(vii) The bracket polynomial is
defined for an unoriented diagram by a sum over ``splittings'':
\eqn\permsplit{
\left<L\right>=\sum_s A^{a(s)-b(s)} (-A^2-A^{-2})^{\# s-1}
}
where the splitting $s$ is described at each vertex by Fig.~\split;
$a(s)$ and $b(s)$ are
the number of vertices of type $(a)$ and $(b)$, and $\# s$ is the 
number of loops thus created.\foot{It might seem surprising that
loops that have non-trivial homology are not distinguished in Eq.~\permsplit;
this is because homeomorphisms of $\Sigma$ and addition/subtraction of handles
do not preserve the homology class of loops.}
Alternatively, $s$ has a simple description as a permutation: define
\eqn\permsplitb{
s(i)=\cases{
\sigma(i)\qquad\hbox{if $i$ overpasses a vertex of type $(a)$}\cr
\tau(i)  \qquad\hbox{if $i$ overpasses a vertex of type $(b)$}\cr
}
}
for all edges $i\in E$. Note that this induces an orientation of the
loops (which is associated
with the bicoloration of the faces). Then $\# s$ is the number of cycles of $s$.
\fig{Splitting at a vertex.}{%
\pic(4312,976)(1929,-1465){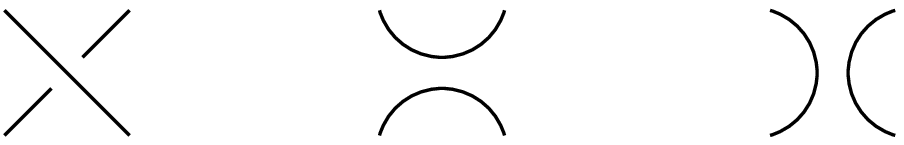}%
\put(3001,-886){$=$}%
\put(3451,-886){$A$}%
\put(4651,-886){$+$}%
\put(5101,-886){$A^{-1}$}%
\put(4000,-1411){$(a)$}%
\put(5800,-1411){$(b)$}%
}{6cm}\figlabel\split

The bracket polynomial is preserved by flypes; however it is
not invariant under Reidemeister move I. It is only
up to multiplication by $-A$ that it is an invariant of unoriented links.

One can get rid of this arbitrary power of $-A$ by introducing
\eqn\permjon{
V=(-A)^{-3t} \left<L\right>
}
where $t=\sum_{1\le i\le c} \ell_{ii}$ is the twisting number of the
link (which is orientation independent).
$V$ is (up to a power of $A$ for multi-component links) the Jones
polynomial in the variable $A=x^{1/4}$.

Furthermore, one can compute colored Jones polyonomials by using cabling, i.e.\ replacing
each string with $k$ parallel strings and then adding extra ``twists'' to keep constant
the linking number of each new string with the original one (e.g.\ keep it zero). We skip
the details of the implementation; let us simply note that the
computation time of the $k$-th cabling roughly grows
like $c_{2k}^{\ n}$ where $c_{2k}={(4k)!\over (2k)!(2k+1)!}$, so that
only $k=2$ can be achieved in a reasonable amount of time.


(viii) The Alexander polynomials are polynomial
invariants of oriented links up to a sign and multiplication by a monomial.
We refer to \refs{\SW,\Saw,\LK,\KR,\BF}
for details. The (extended, multi-variable) Alexander module
is defined by its generators, the edges of the diagram, and local
linear relations at each vertex, see Fig. \alexrule. 
They are very simple to build in terms
of the permutation $\sigma$ once an orientation has been fixed. 
The $0^{\rm th}$ polynomial (which vanishes for classical links)
is simply the determinant of the matrix of relations.
Further polynomial invariants are obtained as g.c.d.\ of minors. 

\fig{The rules defining (1) the Alexander module and (2) the group of a link.}%
{\pic(4800,2160)(901,-2236){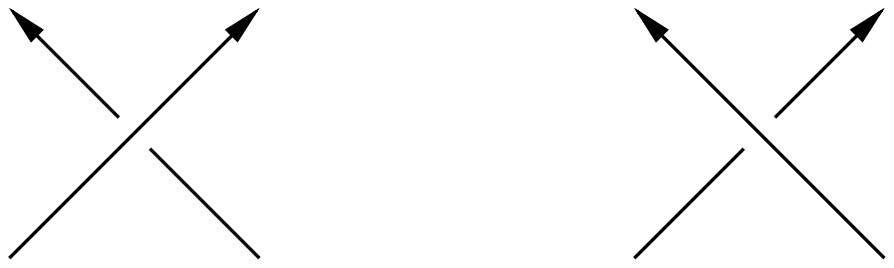}%
\put(4376,-2236){a}%
\put(2726,-2236){a}%
\put(5701,-2236){b}%
\put(1476,-2236){b}%
\put(-300,-386){(1)}%
\put(301,-386){$t_{\rm b}\, {\rm a}+(1-st_{\rm a})\,{\rm b}\qquad s\,{\rm b}$}%
\put(-300,-761){(2)}%
\put(1501,-761){${\rm b} {\rm a} {\rm b}^{-1}\qquad {\rm b}$}%
\put(4026,-386){$s^{-1}\, {\rm b}\qquad t_{\rm b}^{-1}\, {\rm a}+(1-s^{-1}t_{\rm a}^{-1})\,{\rm b}$}%
\put(4426,-761){${\rm b}\qquad {\rm b}^{-1} {\rm a} {\rm b}$}%
}
{6cm}\figlabel\alexrule

(ix) Closely related is the group $\pi$ of a virtual link which is the 
generalization
to virtual links of the fundamental group of the complement of a link; 
the relations at each vertex
are described on Fig.~\alexrule.
It is an invariant (up to isomorphism) of unoriented links (the orientation
only fixes the presentation). 
In practice it is not a simple task (and not easy to implement by computer)
to decide if two groups given by generators and relations are isomorphic, 
and one uses as invariants
the number of morphisms of $\pi$ into given finite groups $\Gamma$.
Unfortunately this is only doable for $\Gamma$ of small order, which only
uncovers a small part of the structure of $\pi$.


\subsec{Generalization to tangles}
\noindent
The appropriate way to consider a (four-legged) tangle 
is a link with a marked rigid vertex.
All that has been done in sections 4.1 and 4.2 can 
therefore be adapted to the case of tangles.

A tangle is now represented by a pair of permutations 
$(\sigma,\tau)$ in which the marked vertex 
is encoded just like an ordinary crossing, except the labels 
must somehow determine uniquely which vertex
is marked: in the implementation we chose the edges overcrossing at
the marked vertex to be
$(2n-2,2n-1)$. The actual tangle is obtained by removing the marked vertex and 
drawing the external 
legs in such a way that $2n-2$ is the lower left line and $2n-1$ 
is the upper right line (and
therefore, $\sigma(2n-1)$ is lower right and $\tau(2n-1)$ is upper left).

The group of relabellings is now limited to those elements of 
$g$ leaving the rigid vertex, 
i.e.\ the corresponding two labels ($2n-2$ and $2n-1$), invariant:
it is a subgroup $G_1$ of $G$ which is isomorphic to ${\cal S}_{n-1} \times
{\Z_2}^{n-1}$. Other operations on the permutations can be performed in the
same spirit, that is by keeping the rigid vertex fixed.

Tangle invariants are generically obtained by pasting an arbitrary
 given tangle to the tangle under consideration
and computing link invariants. In practice one obtains only a finite number of 
independent invariants.
For example, one gets two Jones polynomials, and more generally
$c_{2k}$ for the $k$-cabling, 
by pasting arbitrary arch configurations to the tangle;
five $0^{\rm th}$ Alexander polynomials obtained by setting equal to 0 two of the four 
generators corresponding
to external legs and computing the resulting $(n-1)\times (n-1)$ determinant of the matrix
of relations (there
are ${4\choose2}=6$ possibilities but only 5 are independent due to a bilinear identity satisfied
by the determinants); etc.


\subsec{Results}
\noindent
We have written a program that generates all permutations 
$\sigma\in {\cal S}_n$ up to 
conjugation by elements of $G$ (or $G_1$ for tangles), 
for $n\le 6$. It then selects
connected prime/reduced diagrams, creates
flype equivalence classes, 
and finally sorts them according 
to their invariants and in particular
detects undistinguishable non flype-equivalent diagrams. 
We have provided a sample of the
output on Fig.~\table. 
In particular, the number of morphisms of the group $\pi$ into the
three groups $S_3$, $A_4$ and $A_5$ is listed, but we have
occasionnaly  looked at higher groups. 
To save space, the variable $A$ of the Jones polynomial is called $a$ 
in the tables, and only $0^{\rm th}$ extended 
Alexander polynomials, depending on $c+1$ variables $t_0,\cdots,
t_{c-1}$ and $s$ are listed.
The rest can be found on the web 
({\tt http://ipnweb.in2p3.fr/$\sim$lptms/membres/pzinn/virtlinks}).

{
\def\epsfsize#1#2{0.63#1}
\par\begingroup\parindent=0pt\leftskip=1cm\rightskip=1cm\parindent=0pt
\baselineskip=11pt
\global\advance\figno by 1
\midinsert
\centerline{\epsfbox{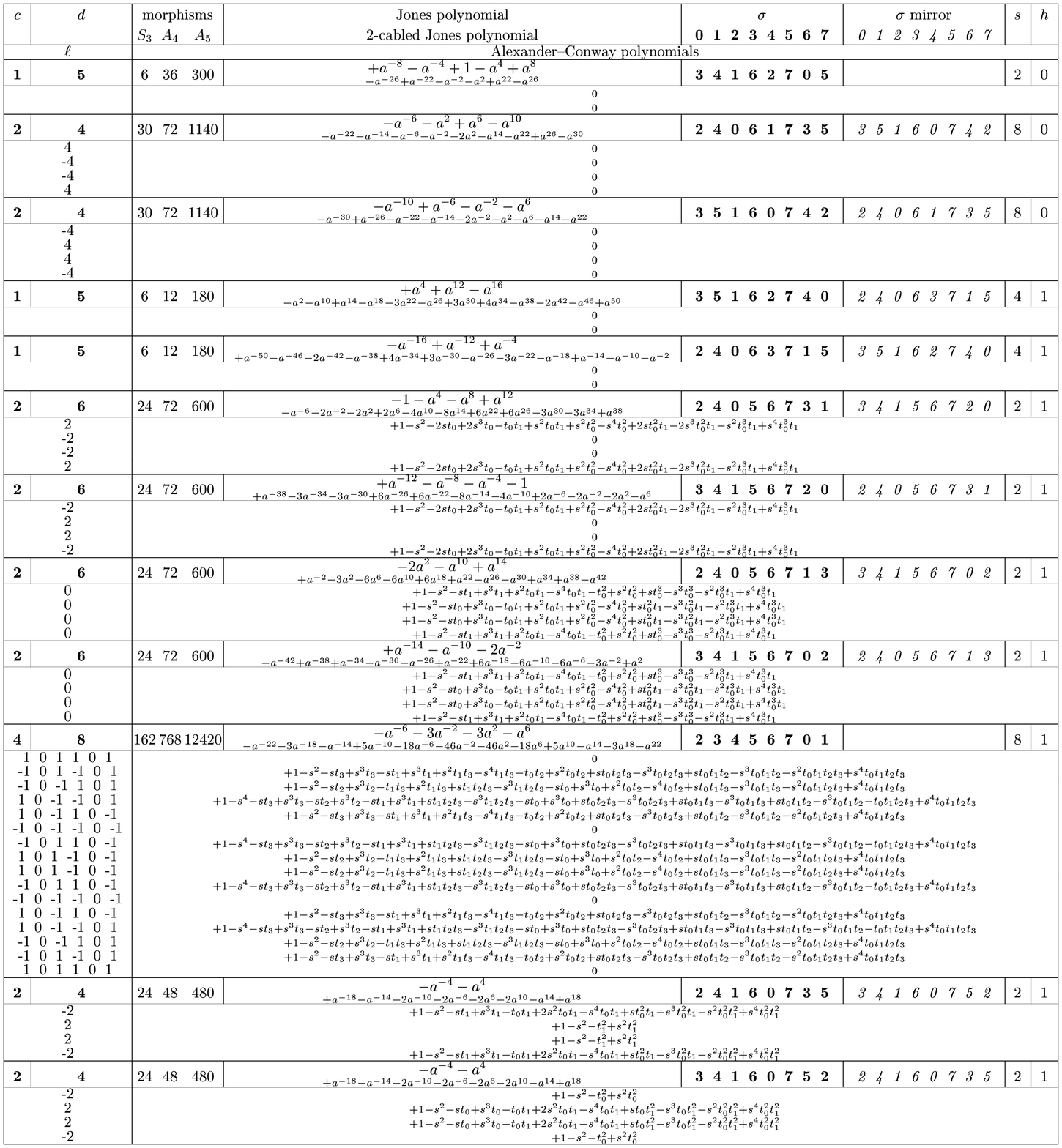}}
\vskip 10pt
{\bf Fig.~\the\figno:} Table of prime alternating links with 4 crossings. 
Mirror images are indicated only
for chiral links. Only $0^{\rm th}$ (extended multi-variable)
Alexander polynomials are listed.
\par
\endinsert
\midinsert
\centerline{\epsfbox{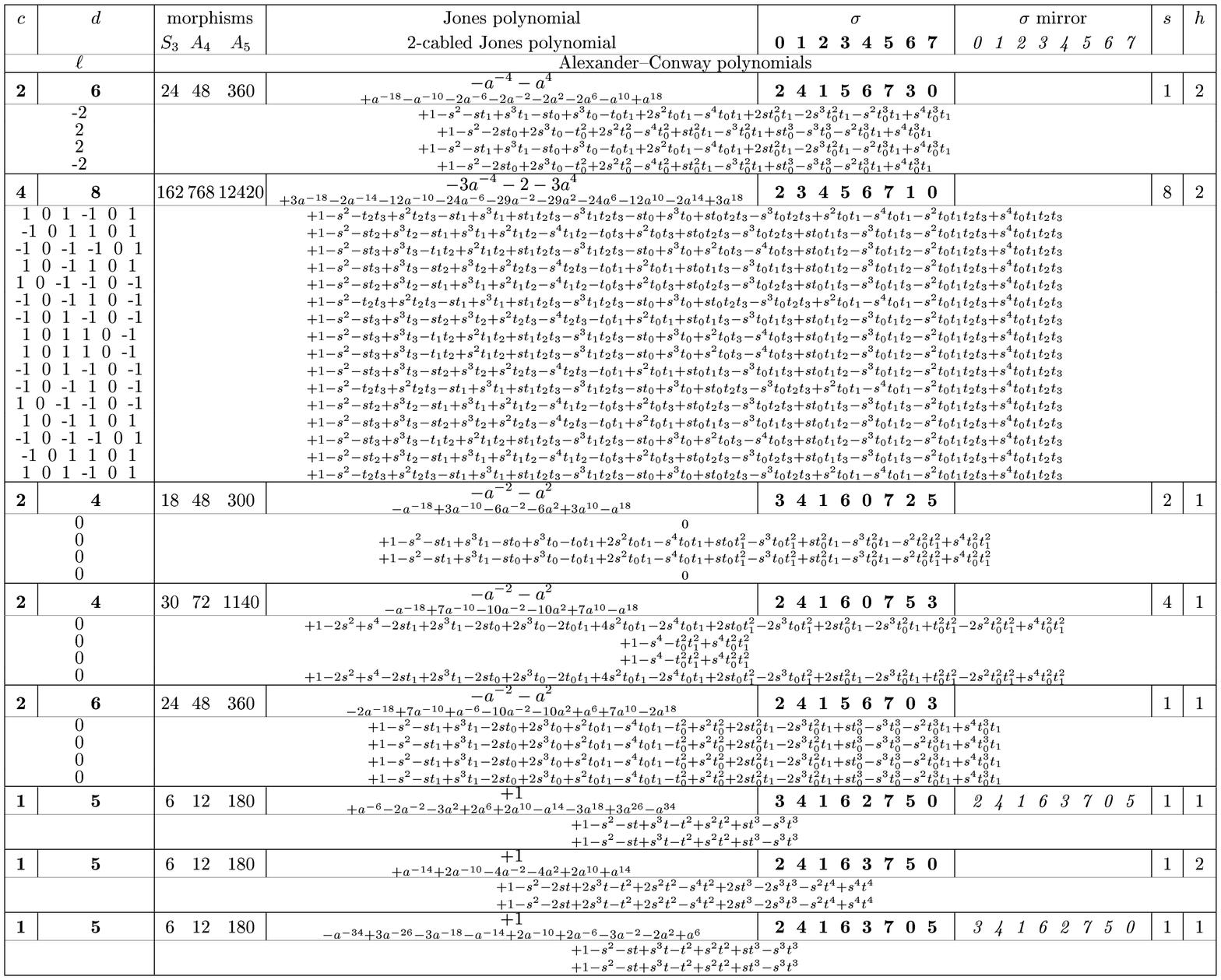}}
\vskip 10pt
{\bf Fig.~\the\figno} (continued)
\endinsert
\endgroup\par
}\figlabel\table

Up to order 4, we have checked that all flype-equivalence classes are distinguished
by the invariants described in section 4.2. Therefore the conjecture holds true
for links at least up to order 4.

When we go to higher orders, a new problem arises: the difficulty in 
distinguishing links obtained from one another by discrete symmetries.


Several discrete operations may be performed within the class
of alternating virtual link diagrams. First, there is the mirror symmetry, 
a reflection in a plane orthogonal to the plane of the figure. This is
a well known operation on classical knots and links, which are called
achiral or chiral depending on whether the mirror image is equivalent 
or not to the original. The same applies to virtual diagrams.
In terms of permutations it corresponds to $\sigma\leftrightarrow\tau$.
Secondly, there is the simultaneous change of all over-
into under-crossings  and vice versa (which corresponds to
$\sigma\to\tau^{-1}$, $\tau\to\sigma^{-1}$). For classical
links, this is not independent 
of the mirror symmetry, as the composition of the two ($\sigma\to\sigma^{-1}$,
$\tau\to\tau^{-1}$), which is
equivalent to a global flip of the diagram, 
i.e.\ a rotation around an axis in the plane of the figure, yields a link 
equivalent to the original: this may be seen by gradually overturning 
the link diagram. 
For virtual diagrams, this is no longer the case:
there is an obstruction to this overturning due to the virtual
crossings, 
and the impossibility of performing the ``forbidden Reidemeister 
move'' of Fig.~\forbid.
Accordingly, there are now some virtual links that are equivalent to 
their flip, and some that are not, the latter appearing at order 5, 
see for example Fig. \renvun. 

In principle, mirror symmetry can be detected by the Jones polynomial since it
corresponds to the transformation $A\to A^{-1}$. 
In fact, even the usual Alexander--Conway
polynomial can distinguish mirror symmetric links in higher genus, 
since it is no longer reciprocal. 
It is important to notice that unlike classical alternating links, 
virtual alternating links do not necessarily 
saturate the bounds on minimum and maximum degrees of their
Jones polynomial (maximum degree minus minimum degree in $A^4$
is less than or equal to $n-h$). For example,
there are many virtual alternating links with trivial Jones
polynomial. Therefore
even detection of mirror symmetry can be tricky in higher genus. 
The situation is worse for the flip symmetry 
since the Jones polynomial (and cabled Jones polynomials)
cannot distinguish flipped images at all. The group of the link or 
the higher Alexander--Conway polynomials
may in some cases distinguish them.

In practice, already at order 5
there are several diagrams which are not related by flypes to their flips
but for which we have not found any invariants to distinguish them.
This is the case for four pairs of links with five crossings, 
namely those of Fig. \renvdeu\ and \renvtroi\ and their mirror images. 
At order 6, there are cases of undistinguishable flips and of 
undistinguishable mirror images.
Based on the experience at genus 0, we believe that
these issues are probably difficult to resolve and
leave them to future work.
\fig{A pair of virtual flipped knots, 
distinguished by their Alexander--Conway polynomial.}
{\epsfbox{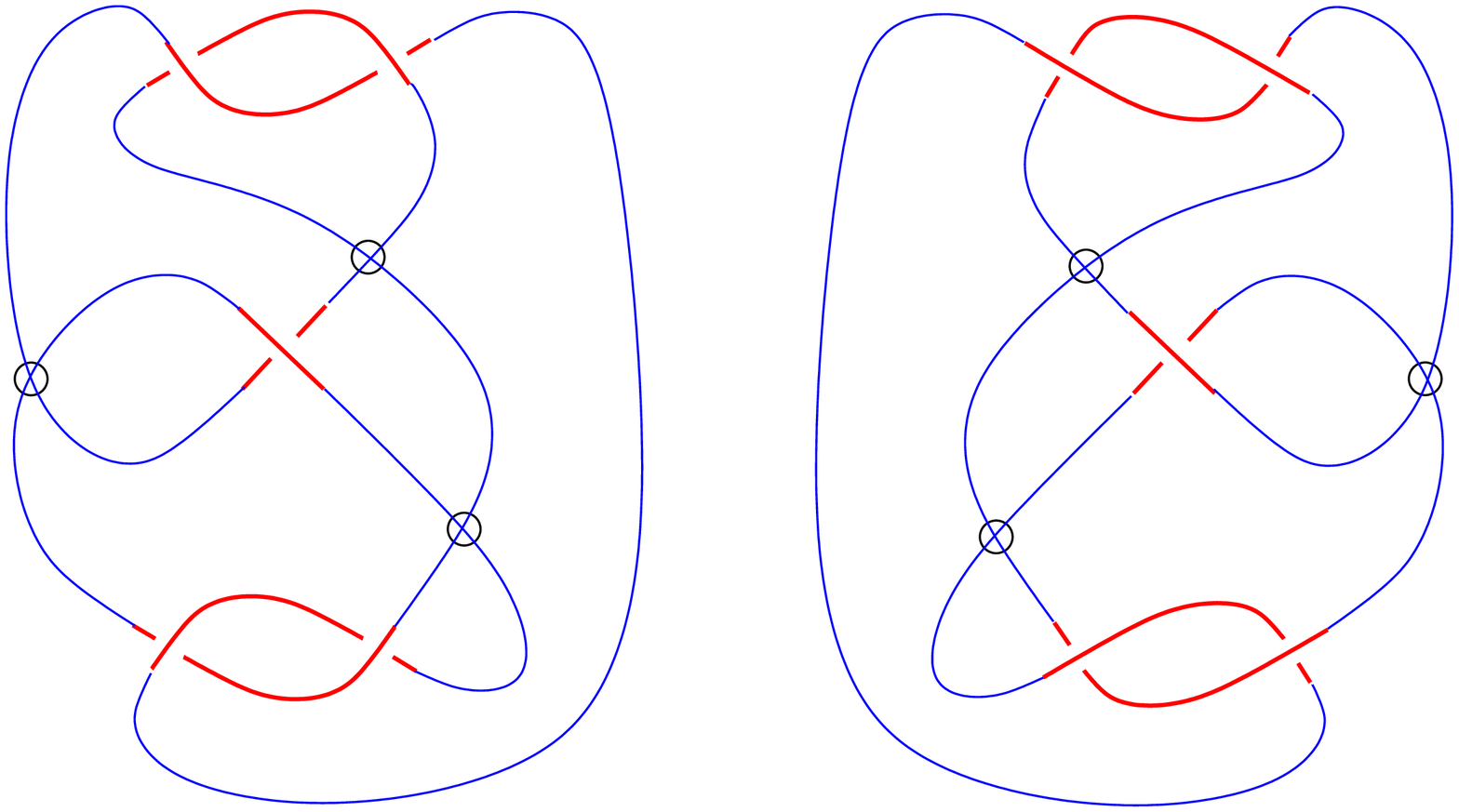}}{7cm}\figlabel\renvun
\fig{A pair of virtual flipped knots of genus 1, 
conjectured to be non equivalent.}
{\epsfbox{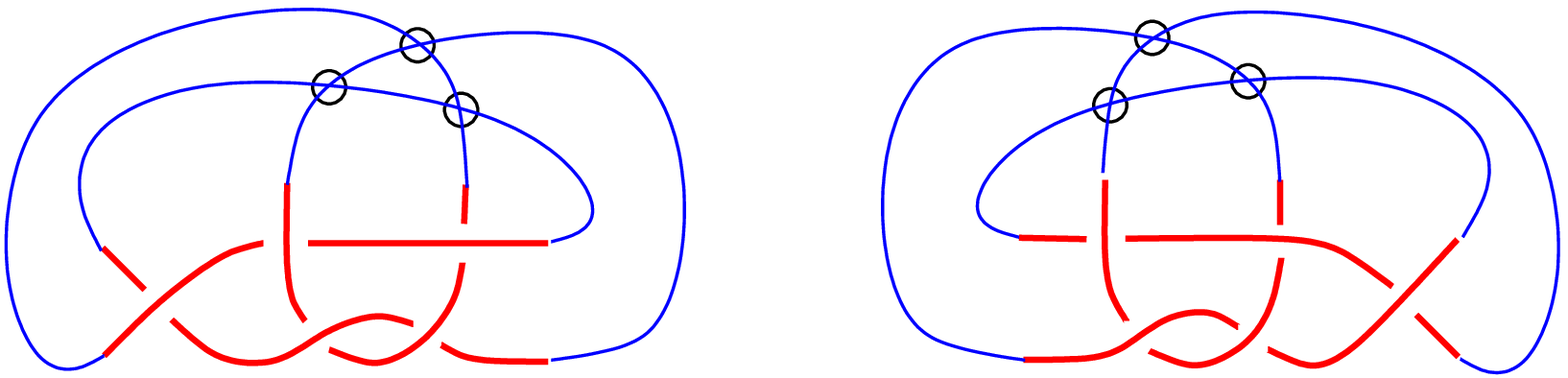}}{8cm}\figlabel\renvdeu
\fig{A pair of virtual flipped knots of genus 2, 
conjectured to be non equivalent.}
{\epsfbox{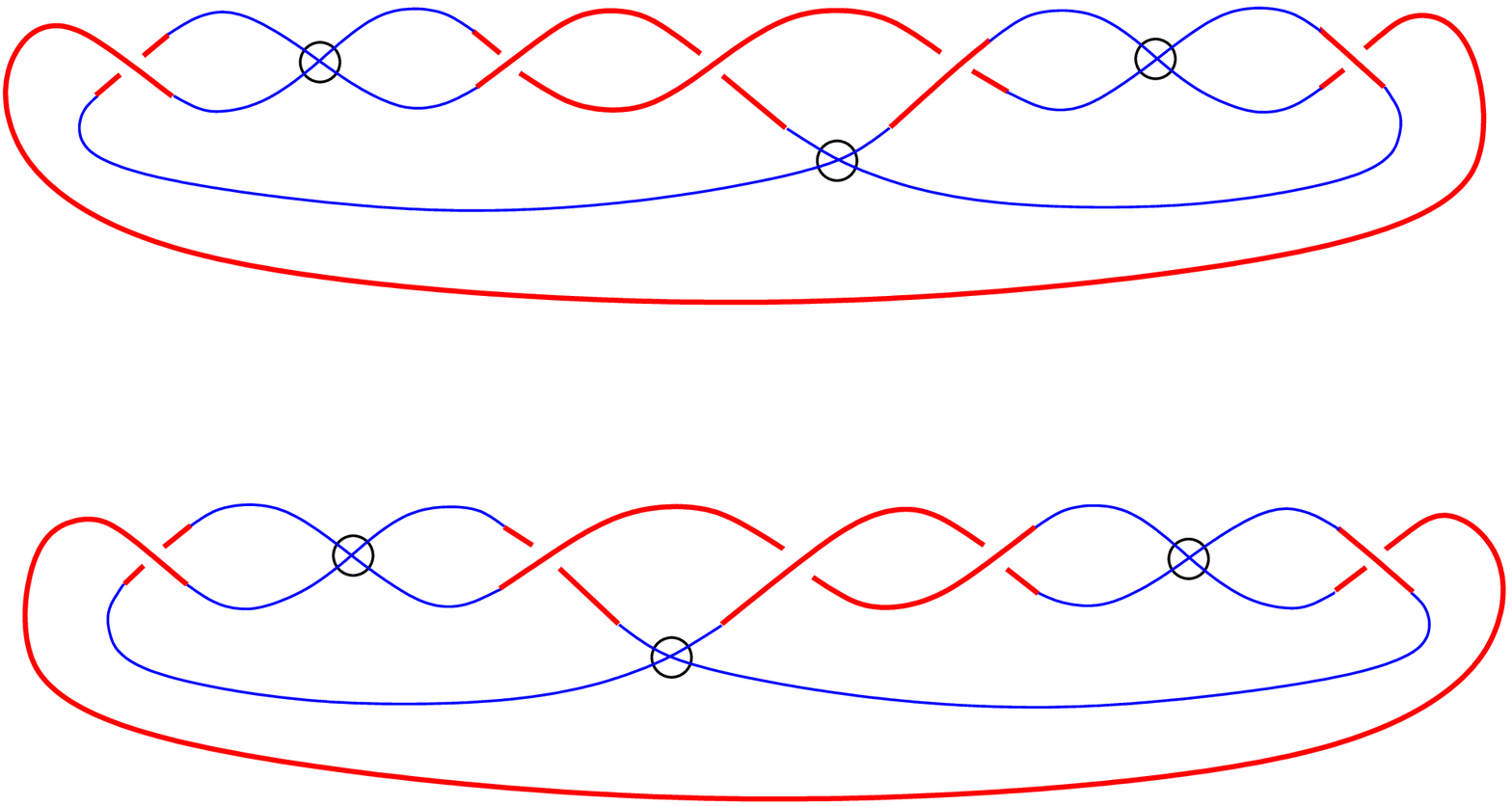}}{7cm}\figlabel\renvtroi

Let us now discuss the case of tangles, which is in fact more important for us
since it is the objects which we enumerate. 
Fortunately the rigid vertex destroys
any possible symmetries and the classification problem becomes easier.
Fig.~\tableb\ shows a very limited sample of our data.
{
\def\epsfsize#1#2{0.63#1}
\fig{Table of prime alternating tangles with 2 crossings. $t$ is the type
of the tangles which encodes how external legs are connected to each other,
according to: NW connected to $t=1$ SE, $t=2$ NE, $t=3$ SW.}{\epsfbox{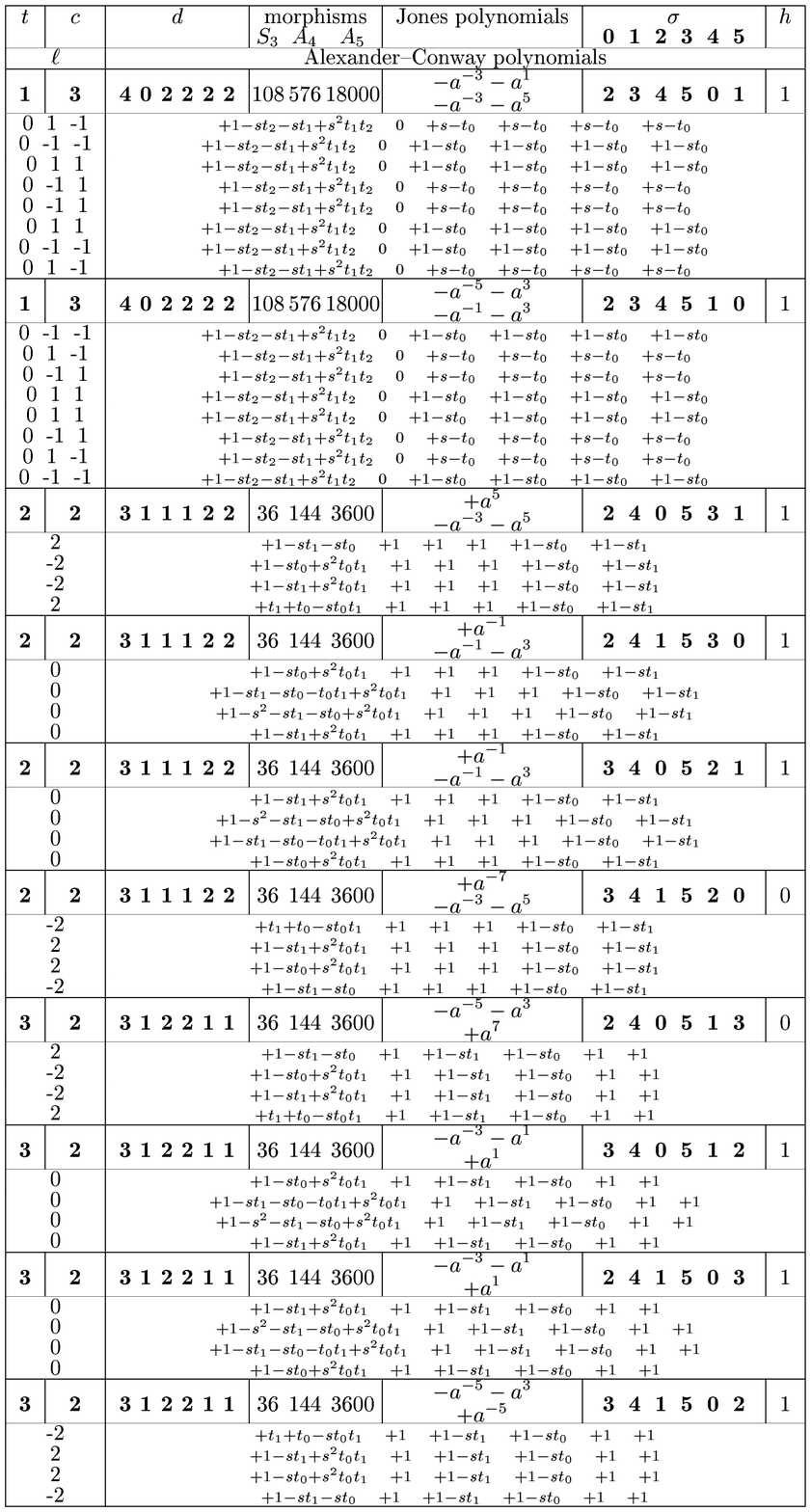}}{0cm}
}\figlabel\tableb

We have compared the 13010 virtual prime reduced
alternating tangle diagrams up to order 5. We have
performed the flype equivalence and checked that the number of tangles
of each genus agrees with Eq. \Gamtild. 
Our program
then allows us to assert that all the flype-equivalence classes
thus obtained, irrespective of genus
and crossing number, are distinct (note that in a few cases
we had to manually feed the computer groups of fairly large order -- up to 432 --
to make it distinguish the corresponding link groups via their morphisms). 
This we consider a strong argument in favor of
our generalized flype conjecture.

\vskip1cm

\bigskip
\centerline{\bf Acknowledgments}\nobreak
It is a pleasure to thank 
L.~Funar, L.~Kauffman, G.~Kuperberg, P.~Vogel for discussions,
and G.~Akemann for providing us with
unpublished data by P.~Adamietz and himself. 
\hbox{J.-B.~Z.}\ is partially supported by the European network 
HPRN-CT-2002-00325.

\listrefs


\appendix{A}{Computation of the lowest order term of genus $h$} 
       
\noindent
According to the description of the $g$ expansion in terms of 
permutations of $S_{2n}$, the leading term with  $n=2h$ reads 
\eqn\Aa{F_{2h}^{(h)}={1\over (2n-1)!!}{1\over 2^{n}\,n!}
\sum_{\Gs,\Gt\in \S_{2n}}   \Gd_{[\Gs],\{2n\}} \Gd_{[\Gt],\{2n\}} 
\Gd_{[\Gs^{-1}\Gt],\{2^{n}\}}}
where the factor ${1\over 2^{n}\,n!}$ comes from the $n$-th order
of the $g$ expansion, and ${1\over (2n-1)!!}$ takes care of the 
remaining relabeling invariance of the permutations. 
Also $[\sigma]$ denotes the class of the permutation $\sigma$. 
One then uses the orthonormalized characters of the $\S_{2n}$ 
symmetric group to represent the conditions on $\Gs$ or $\Gt$
by $\delta_{[\Gs],\{\Ga\}}=
\sum_Y 
{\nu_{\Ga}\over (2n)!}\chi_Y(\Ga)\chi_Y([\Gs])$ 
where the sum runs over Young tableaux 
with $2n$ boxes, and 
$\nu_{\Ga}$ is the number of elements of the 
class $\Ga$: if $\Ga=\{1^{\Ga_1}2^{\Ga_2}\cdots\}$, 
$\nu_{\Ga}=(2n)!/\prod_j(\Ga_j! j^{\Ga_j})$. Thus
\eqnn\Ab
$$\eqalignno{
(4h)! F_{2h}^{(h)}&= \sum_{\Gs,\Gt\in \S_{4h}} 
\({\nu_{\{4h\}}\over (4h)!}\)^2
\({\nu_{\{2^{2h}\}}\over (4h)!}\) \cr & \hskip2cm \times \sum_{Y_1,Y_2,Y_3}
\chi_{Y_1}([\Gs])\chi_{Y_1}(\{4h\})
\chi_{Y_2}([\Gt])\chi_{Y_2}(\{4h\})\chi_{Y_3}([\Gs^{-1}\Gt])\chi_{Y_1}(\{2^{2h}\})
\cr
&= {\nu_{\{4h\}}^2 \nu_{\{2^{2h}\}}\over (4h)! }
\sum_{Y}{1\over d_Y} \(\chi_{Y}(\{4h\})\)^2  \chi_{Y}(\{2^{2h}\})\ . & 
\Ab
}$$ 
The characters $\chi_{Y}(\{2n\})$ for the one-cycle class receive 
contributions only from the hook Young tableaux  \ 
\def\tv{\vrule height 4pt depth 2pt} 
\def\th{\vrule height 0.4pt width 0.7em} 
\def\ca{\hfill{\ }\hfill} 
\hskip-10mm
\setbox111=\hbox{$\vbox{\offinterlineskip
\+          &&   &\th&\th&\th&\th&\th&\th&\th&\th&\th&\cr
\+
&\tv&\ca&\tv\ca&\tv\ca&\tv\ca&\tv\ca&\tv\ca&\tv\ca&\tv\ca&\tv\ca&\tv\ca&
\tv&\cr
\+       &  &&\th&\th&\th&\th&\th&\th&\th&\th&\th&\cr
}$}
\setbox12=\hbox{$\vcenter{\offinterlineskip
\+       &\tv&\ca&\ca&\tv&   &   &   &   &   &\cr
\hrule
\+       &\tv&\ca&\ca&\tv&   &   &   &   &   &\cr
\hrule
\+ \cr
\+        & \tv&\ca&\ca&\tv&   &   &   &   &   & \cr
\hrule }$}
\setbox22=\hbox{$\left.\vbox to \ht12{}\right\}$}
\setbox33=\hbox{$s$}
\setbox10=\hbox to 6.2em {\downbracefill}      
$Y_s=\vcenter{\vbox{\offinterlineskip\halign{
#& \qquad #& \hfill # \cr
\hbox to 6.5em{\hfill\quad$ 2n-s$ \quad\hfill}\cr
\noalign{\vskip 2mm}
\box10 \cr
\noalign{\vskip 2mm}
\copy111 \cr
\copy12\copy22\box33 \cr  }}}$
\quad for which 
\def\Ys{{{}\atop \scriptstyle{Y}_s}}
$$\eqalign{ \chi_{\Ys}(\{2n\})&=(-1)^s, \cr  
\chi_{\Ys}(\{2^{n}\}&=
(-1)^{\lfloor{s+1\over 2}\rfloor} {n-1 \choose \lfloor{s\over 2}\rfloor}\cr
d_{\Ys}& ={(2n-1)!\over s!(2n-s-1)!}\ .}$$
The summation in \Ab\ is thus reduced to a sum over $s=0,\cdots, 4h-1$. 
After some algebra, one finds the result \Fmin.

The first subleading term $F_{2h+1}^{(h)}$ is given in a similar way 
by  
\eqn\Ac{F_{2h+1}^{(h)}={1\over (2n-1)!!}{2 \over 2^{n}\,n!}
\sum_{\Gs,\Gt\in \S_{2n}} \sum_{p=1}^{n}  \Gd_{[\Gs],\{2n\}} 
\Gd_{[\Gt],\{p,2n-p\}} 
\Gd_{[\Gs^{-1}\Gt],\{2^{n}\}}}
with now $n=2h+1$, 
since $F=3$ implies that either $\#\sigma=1$, $\#\tau=2$ or the converse.
The same method as above applies, again only hook Young tableaux 
contribute, and the only additional piece of information required is
$$  \chi_{\Ys}(\{2n-p,p\})=\cases{(-1)^s & if $ s+1\le p\le 2n-s-1$, \cr
(-1)^{s+1} & if $ 2n-s\le p\le s$ \cr 0   & otherwise\cr
}\ ,$$
as we learn from the Murnaghan-Natayama formula. 
A little algebra then leads to the result \Fnlo.

\bye